\pdfoutput=1
\documentclass[MNSC]{informs3nothing} 
\RequirePackage[OT1]{fontenc}
\RequirePackage{
amsmath,
times,
graphicx,
amsfonts,
amscd,
amssymb,
algorithm,
algorithmic,
bbm,
multicol
}
\OneAndAHalfSpacedXI

\usepackage[sort&compress,comma,authoryear]{natbib}
 \bibpunct[, ]{(}{)}{,}{a}{}{,}%
 \def\bibfont{\small}%
 \def\bibsep{\smallskipamount}%
 \def\bibhang{24pt}%
\TheoremsNumberedThrough     
\ECRepeatTheorems

\EquationsNumberedThrough    

\MANUSCRIPTNO{} 

\newcommand{\paranth}[1]{\left(#1\right)}
\newcommand{\bracket}[1]{\left[#1\right]}
\newcommand{\curly}[1]{\left\{#1\right\}}

\numberwithin{equation}{section}
\theoremstyle{plain}
\newtheorem{thm}{Theorem}[section]

\newtheorem{lem}{Lemma}

\begin{document}


\RUNAUTHOR{Hunter, Saini, and Zaman}

\RUNTITLE{Winning}

\TITLE{Picking Winners: A Data Driven Approach to Evaluating the Quality of Startup Companies }

\ARTICLEAUTHORS{%
\AUTHOR{David Scott Hunter}
\AFF{Operations Research Center, Massachusetts
Institute of Technology, Cambridge, MA 02139, \EMAIL{dshunter@mit.edu}} 

\AUTHOR{Ajay Saini}
\AFF{Department of Electrical Engineering and Computer Science, Massachusetts
Institute of Technology, Cambridge, MA 02139, \EMAIL{ajays235@mit.edu}} 

\AUTHOR{Tauhid Zaman}
\AFF{Department of Operations Management, Sloan School of Management, Massachusetts
Institute of Technology, Cambridge, MA 02139, \EMAIL{zlisto@mit.edu}}
} 

\ABSTRACT{We consider the problem of evaluating the quality of startup companies.     This can be quite challenging  due to the rarity of successful startup companies and the complexity of factors which impact such success.  In this work we collect  data on tens of thousands of startup companies, their performance, the backgrounds of their founders, and their investors.  We develop a novel model for the success of a startup company based on the first passage time of a Brownian motion.  The drift and diffusion of the Brownian motion associated with a startup company are a function of features  based its sector, founders, and initial investors.  All features are calculated using our massive dataset. Using a Bayesian approach, we are able to obtain quantitative insights about the features of successful startup companies from our model.  
	
To test the performance of our model, we use it to build a portfolio of companies  where the goal is to maximize the probability of having at least one company achieve an exit (IPO or acquisition), which we refer to as winning. This \emph{picking winners} framework is very general and can be used to model many problems with low probability, high reward outcomes, such as pharmaceutical companies choosing drugs to develop or studios selecting movies to produce.  We frame the construction of a picking winners portfolio as a combinatorial optimization problem and show that a greedy solution has strong performance guarantees.  We apply the picking winners framework to the problem of choosing a portfolio of startup companies. Using our model for the exit probabilities, we are able to construct out of sample portfolios which achieve exit rates as high as 60\%, which is nearly double that of top venture capital firms. 
	
}

 \HISTORY{}

\maketitle

\section{Introduction}\label{sec:intro}
 An important problem in entrepreneurship is evaluating the quality of a startup company.  This problem is known to be extremely challenging for a variety of reasons. First, successful companies are rare, so there are not many to study in order to discern patterns of success. Second,  at the time a startup company is founded, there is not very much data available.  Typically one only knows basic information about the company's founders, the sector or market in which it operates, and the identity of initial investors.  It is not a priori clear if this raw data is sufficient to allow one to measure a startup company's quality or predict whether or not it will succeed.  Third, it is not obvious how to model the evolution of a startup company and how to use this model to measure its quality.  Fourth, one needs a principled approach to test the validity of any startup company model or quality measure.

The problem of evaluating startup company quality is faced by venture capital firms.  These firms spend a large amount of time and resources screening and selecting startup company deals \citep{metrick2010venture,da2011survey,sahlman1990structure}.  In particular, firms take a hands-on approach to investing by evaluating the attractiveness and risk behind every investment, considering factors that include relative market size, strategy, technology, customer adoption, competition, and the quality and experience of the management team. The investment screening process is oftentimes a thorough and intensive one that can take several months.  Venture capital firms could greatly improve their investment process if they had a quantitative model which could perform a preliminary evaluation of the quality of a large number of potential deals.  They could use such a model  to prioritize and more efficiently allocate their human resources for evaluating different deals.
 
 While venture capital firms still make up a large majority of the investors in early-stage startups, with the creation and adoption of the Jumpstart Our Business Startups (JOBS) Act \citep{congress2012jumpstart}, early-stage investment opportunities are now available to the everyday investor in the United States. Before this policy change, only certain accredited investors were allowed to take equity stakes in most private companies. With the adoption of this policy, there are now many companies such as FlashFunders and NextSeed \citep{FlashFunders,NextSeed} that assist the everyday investor in finding and participating in startup investment opportunities. While these sites provide some relevant information about the startup companies of interest, 
 users of these sites do not have direct access to the company in the same way that a venture capital firm does. Thus, oftentimes they do not have the information needed to execute a thorough and intensive selection process.  One expects that these investment opportunities will have a lower success-rate than that of venture capital backed startups.  This means that there is a need for quantitative measures of startup company quality which can be developed from publicly available data.  Such measures can serve to level the playing field and empower the everyday investor.

It is useful to take an operational perspective for validating a model for startup company quality.  Such a model would be part of the process used by venture capital firms or other investors to build portfolios of startup companies.  Therefore, the operational value of a model is its ability to produce successful portfolios.  In typical portfolios of startup companies, a few strong performers are responsible for most of the returns \citep{angelblog}. Startups  can produce excessively high returns if they exit, which means they are either acquired or they have an initial public offering (IPO).  For instance, Peter Thiel invested \$500,000 in Facebook for 10\% of the company in 2004, and in 2012 he sold these shares for \$1 billion, resulting in a return of 2,000 times his initial investment \citep{thiel_facebook}.  

In general, an investor in a startup company will lose money unless the company exits, which we refer to as winning.  Because only a few winners are needed for a portfolio to be profitable, one way to build a portfolio is to  pick companies that maximize the probability of at least one winner.    We refer to this as the \emph{picking winners problem}. In addition to building portfolios of startup companies, this framework can be used to model a variety of problems which have a low probability, high reward structure.  For instance, most new drugs developed by pharmaceutical companies fail.  However, with a small probability some of the drugs are successful, and having at least one winning drug can be enough for a company to be successful.  For another example, consider a studio that has to select a set of movies to produce.  The studio will be successful if they can produce one blockbuster. In these examples, one approach to the problem is to select a portfolio of items (drugs, movies, startup companies) to maximize the probability that at least one of them has exceptional performance, or ``wins''.

\subsection{Our Contributions}
In this work we present a data driven approach to evaluate the quality of startup companies.  Our results are especially useful to venture capital firms which may have to analyze thousands of startup companies before making investment decisions.  We provide a way to quantify the quality of a startup company.  Our work involves a large, comprehensive dataset on startup companies, a novel model for their evolution, and a new approach to validate our model. 

Our first contribution is the collection and analysis of a massive dataset for several thousand companies.  This data consists of information on the companies' funding round dates, investors, and detailed professional and educational history of the founders.
Exploratory analysis of this data reveals interesting properties about the progression of the funding rounds of startup companies and factors that impact their success.

Our second contribution is a  model for the evolution of a startup company's funding rounds.  This models describes a startup company using a Brownian motion with a company dependent drift and diffusion.  The 
first passage times of the Brownian motion at different  levels correspond to the startup company receiving a new round of funding, and the highest
level corresponds to the company exiting.  One useful property of our model is that it naturally incorporates the censoring that occurs due to limited observation times.  
We estimate the model parameters using a Bayesian approach.  This allows us to perform a statistical analysis  of features that impact company success.

Our final contribution is a validation of our model with picking winners portfolios.  We build such portfolios for future years using model estimation results on past data.
We find that our portfolios achieve exit rates as high as 60\%.  The performance of our portfolios exceeds that of top venture capital firms.  This not only provides evidence supporting our model, but it also demonstrates its potential operational utility for investment decisions.

The remainder of the paper is structured as follows.  We begin with a literature review in Section \ref{sec:literature_review}.    We present an exploratory analysis of our startup company data  in Section \ref{sec:data}. In Section \ref{sec:model} we  introduce our Brownian motion model for startup funding.  We develop Bayesian formulations of our core model in Section \ref{sec:bayesmodelsection}. In Section \ref{sec:estimation} we discuss the estimation of our models. We study the insights gained from the  estimation results in Section \ref{sec:estimationresults}.  In Section \ref{sec:problem} we formulate the picking winners problem and discuss some of its theoretical properties and relationship with submodular optimization and  log-optimal portfolios.  We present the performance of picking winners portfolios based on our models in Section \ref{sec:portfoliosection}. We conclude in Section \ref{sec:conclusion}. All proofs as well as more detailed information about the data, model formulation, and model estimation are included in the appendix.

\section{Literature Review}\label{sec:literature_review}

There have been several studies that take a quantitative approach to determining what factors are correlated with startup company success.  It is argued in \cite{porter2001innovation} that geographic location is a important factor for success in innovation activity.  The work in \citep{guzman2015silicon} and \citep{guzman2015nowcasting}  use a logit model to predict startup success based on features such as company name, patents, and registration, and then use the model to assess the entrepreneurial quality of geographic locations.   In \citep{azoulay2018age} it is found that that successful entrepreneurs are generally older.
\cite{eesley2014contingent} has found that founding team composition has an effect on the startup's ultimate performance . In particular, this study found that more diverse teams tend to exhibit higher performance, where higher performance is measured by exit rate. Additionally, researchers have considered how startup success correlates with how connected the founders are within a social network \citep{nann2010power}.
Another study \citep{gloor2011analyzing} considers how startup entrepreneurs' e-mail traffic and social network connectedness correlate with startup success. In \cite{marmer2011startup}, the researchers studied how the founders of Silicon Valley startups influence a company's succes and found that the founders' ability to learn and adapt to new situations is most important to company success.  In \citep{xiang2012supervised} the authors use topic modeling on news articles to predict mergers and acquisitions.  \cite{nanda2017persistent} find that with venture capital firms, successful investments increase the likelihood of success in future investments.

The question of building portfolios of startup companies was studied in \citep{bhakdi2013quantitative}, where a simulation-based approach was used to properly structure a portfolio of early-stage startups to adjust for market volatility. 
In the private sector,  several venture capital firms have been adopting analytical tools to assist in the investment process. Google Ventures \citep{GV}, the venture capital arm of Alphabet, Inc., is known to use quantitative algorithms to help with investment decisions \citep{Fortune}. Social Capital \citep{socialcapital} is a venture capital firm that has a service called capital-as-a-service which makes investment decisions based on data about a company \citep{socialcapitalmedium}. Correlation Ventures \citep{CorrelationVentures},  a venture capital firm with over \$350 million under management, uses a predictive model to make startup company investment decisions. It ask startups to submit five planning, financial, and legal documents, it takes only two weeks to make an investment decision, and it never takes administrative board seats. Interestingly enough, Correlation Ventures only takes co-investment opportunities and relies on some vetting by a primary investor \citep{CorrelationVenturesReview}.

There is a large body of research related to the  picking winners problem.  There is a natural relationship between this problem and a variety of set covering problems.   There is the problem of placing a fixed number of sensors in a region to detect some signal as fast as possible.  This has been studied for water networks \citep{leskovec2007cost} and general spatial regions \citep{guestrin2005near}.  There is also a problem in information retrieval of selecting a set of documents from a large corpus to cover a topic of interest \citep{chen2006less, el2009turning, swaminathan2009essential, agrawal2009diversifying}.  Another related problem is selecting a set of users in a social network to seed with a piece of information in order to maximize how many people the message reaches under a social diffusion process \citep{kempe2003maximizing}.  Another way to view this problem is that one wants the message to reach a certain target user in the network, but one does not know precisely who this user is, so one tries to cover as much of the network as possible.  One can view the sensors, documents, or seed users as items in a portfolio and winning means one of the sensors detects the signal, one of the documents covers the relevant topic, or one of the seed users is able to spread the message to the unknown target user.  

The similarity between picking winners and these covering problems has important implications.  The objective functions in these covering problems are submodular and maximizing them is known to be NP-hard \citep{karp1998nphard}.  The same holds for the picking winners problem.  The submodular property is important because it means that one has  a lower bound on the performance of a greedy solution \citep{nemhauser1978analysis}.  This fact, plus the simplicity with which one can calculate greedy solutions, motivates us to use a greedy approach to pick winners.  

There are also interesting connections between the picking winners problem and traditional portfolio construction methods.
For instance, if the items have binary returns, meaning the return is either very high or very low,
 then there are striking similarities between the optimal portfolio in the picking 
winners problem and what is known 
as the log-optimal portfolio \citep{cover1991universal}.  In fact, we show in Section \ref{sec:problem} that when there is independence the portfolios
are equivalent, and under certain conditions, the picking winners portfolio produces a strong solution to the log-optimal portfolio selection problem.


\section{Data Analysis}\label{sec:data}

We collected a large amount of relevant data regarding startup company performance and features.
For each company, there were two categories of data we acquired: 
the time it reaches different funding rounds 
and features relating to its sector, investors, and leadership.  We obtained our data from
three different sources.  The first is Crunchbase \citep{crunchbase}, a public
database of startup companies.  The information in Crunchbase is provided by
public users, subject to approval by a moderator \citep{crunchbase_contributors}.
The second is a privately maintained database known as Pitchbook that also collects 
information on startup companies \citep{pitchbook}.   The two databases
combined provide us with data for over 83,000 companies founded between 
1981 and 2016. These databases give us information about when a company receives a funding round or achieves an exit, and which investment groups are involved in a particular funding round.  Additionally, these databases provide us with information on over 558,000 company employees. In particular, we have information regarding  which companies a person has worked for, their roles within these companies, and their employment timeline. To build on this person-level information, we also use LinkedIn, a business and employment-focused
 social network \citep{linkedin}.  Including LinkedIn data gave us  complete educational and occupational histories for the employees.

\subsection{Funding Rounds Data}
A startup company typically receives funding in a sequence of rounds.  
The initial funding is known as the seed round.  
After this, the company can receive funding in a series  of `alphabet rounds' which are called series A, series B, etc.  The alphabet rounds typically do not go beyond series F. Additionally, a company can reach an exit, which is when a startup company is either acquired or has an IPO.   Crunchbase provided the funding round and exit dates for each startup company, while the 
Pitchbook data that we used only provided the founding date.  Therefore, our base set
of companies was dictated by Crunchbase. We used Pitchbook to resolve
founding date errors in Crunchbase.  For instance,
there were some companies in Crunchbase where the listed founding date
 occurred after later funding rounds.  In these cases, we ignored the Crunchbase founding
date and used the Pitchbook date.

\begin{figure}
	\includegraphics[scale=1]{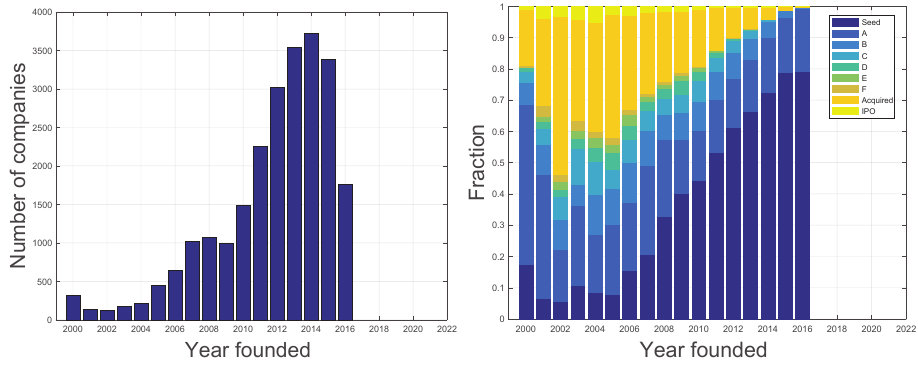}
	\caption{(left) Histogram of the founding year of the startup companies in our dataset.
	(right)  Distribution of the maximum funding round achieved up to 2016 broken down by company founding year.}
	\label{fig:milestones}
\end{figure}

For our analysis we only use companies founded in 2000 or later.  We use this cutoff because 
before this year the databases did not contain very many unsuccessful companies, creating a biased dataset. Furthermore, for our analysis we are particularly interested in measuring the likelihood of a company in an early round eventually reaching an exit. Therefore, we only consider startup companies where we have reliable information on their  seed or series A funding rounds. In particular, if we do not have information on when a startup company reached one of these early rounds, then we omit them from our analysis. Additionally, we chose to focus on startup comanies that are founded in the United States. Using 2000 as a cutoff year and only considering American companies with reliable information for their early rounds left us with 
approximately 24,000 companies founded between 2000 and 2016.  We plot the number
of companies founded in each year in Figure \ref{fig:milestones}.  As can be seen, 
there are a few companies in the early years, but by 2011 the number
of companies increases by a large amount.  This increase is most likely
due to the fact that Crunchbase was founded in 2011 and after this date
many companies began entering their information into the database.  The drop in the number of companies in 2016 is due
to the fact that we collected our data in the middle of 2016.

We first wanted to understand the distribution of the maximum funding round achieved (as of 2016) for these companies.  We plot
in  Figure \ref{fig:milestones} this distribution broken down by year.  We observe a few interesting
properties from this figure.  First, the average fraction of companies founded in a given
year that have an IPO is typically a few percent.  For acquisitions, this value goes up to 21\%, but as
can be seen, the fraction of acquisitions changes a great deal over the years.  After
2011, the fraction of acquisitions falls below 14\% and decreases each subsequent year.
The decrease is most likely due to data censoring, as companies founded in these later years
may not have had enough time to exit.  From 2000 to 2002 the fraction of acquisitions
actually increases.  This may be due to the small number of companies in our dataset and sampling bias
leading to an under-reporting of less successful companies.

\begin{figure}
	\includegraphics[scale=1]{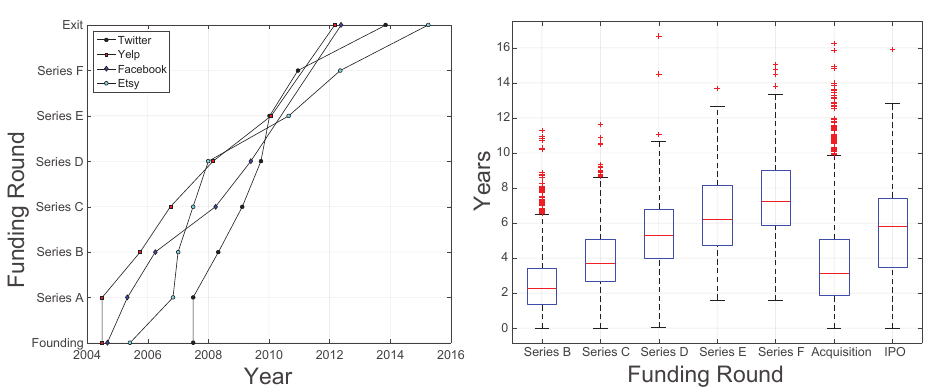}
	\caption{(left) Plot of the evolution of funding rounds for different startup companies.
	(right)  Boxplot of the time to reach different maximum funding rounds before 2016.}
	\label{fig:milestones_time}
\end{figure}

We next wish to understand the temporal dynamics of the funding rounds.
Figure \ref{fig:milestones_time} shows the evolution of funding rounds
for different well known companies that achieved IPOs.
As can be seen, the rate of achieving an IPO fluctuates between companies.
The boxplot in Figure \ref{fig:milestones_time} shows the distribution of the
time to hit the different rounds.  From here it can be seen that IPOs happen
after about six years, but acquisitions typically take three years.   The time between
the other funding rounds (series B, series C, etc.) is about a year.  The typical
variation in these times across companies is about one to two years.  Furthermore, 
we can see that most of the acquisitions happen around the same time that most companies
are at series C funding. Beyond this point, there are fewer companies that get acquired, thus
indicating that as a company gets older and progresses past series C, it's exit will more likely be
an IPO rather than an acquisition if it ever exits.

\subsection{Sector, Investor, and Leadership Data and Features}\label{sec:company_data}
We now describe the data and features that we used in our model. We were able to build a wide variety of features for our model, ranging from simple features such as a company's sector, to more complex features relating to the similarity of the academic and occupational backgrounds of the startup founders. The features for each company are constructed using only information known at the time the company received it's first funding round, which we refer to as the \emph{baseline date} of the company. We now describe these features in more detail.

\textbf{Sector Data and Features.}
The majority of the features that were used in our model relate to the sector that a startup company was involved in. In particular, from Crunchbase we were able to obtain sector labels for each company, with some of the companies belonging to multiple sectors. While Crunchbase has informative labels for many different features, we chose to include binary indicators for a subset of them. We made a point to include a wide-variety of sectors in our model, ranging from \emph{fashion} to \emph{artificial intelligence}. We provide a complete list of the sectors that we used in Appendix \ref{sec:sector_names}.

\textbf{Investor Data and Features.}
From the Crunchbase data, we are able to construct a dynamic network of investors and companies, such that there is a node for every investor and company that exists at the particular time of interest, and there is an edge connecting an investor to a company if the investor has participated in a funding round for that company before the particular time of interest. We construct this dynamic network using all of our available data -- meaning that we consider roughly 83,000 companies and 48,000 investors. We derive features for an individual company based on this dynamic network at the time when the company had its first funding round. Recall that we omit companies without reliable time information for their seed or series A funding rounds. Therefore, for a particular company we consider the dynamic network of investors at the time of the earliest of these rounds.  

Using this dynamic network, we construct a feature that we call
the \emph{investor neighborhood}.  For a company $i$ having an earliest funding date of $t_i$, 
the value of this feature is the number of startup companies in existence before year $t_i$ that share 
at least one investor in common with company $i$.  We then normalize this value
by the total number of companies founded before $t_i$.  This feature measures the
relative reach of the company's investors.

Another feature that is derived from this dynamic network is the \emph{maximum IPO fraction}.  For each initial investor $j$ connected to company $i$, 
we define $f_j$ as the fraction of companies connected to $j$ at $t_i$ that also had an IPO before $t_i$.
The feature value is then the maximum value of $f_j$ amongst all initial investors in $i$.  

A related feature we define is called \emph{maximum acquisition fraction}.  This feature is identical
to \emph{maximum IPO fraction}, except we use the fraction of companies that were acquired rather than the fraction
that had an IPO.  Both \emph{maximum IPO fraction} and \emph{maximum acquisition fraction} are measures of the success 
rate of a company's initial investors.

\begin{figure}
	\begin{center}
	\includegraphics[scale=.35]{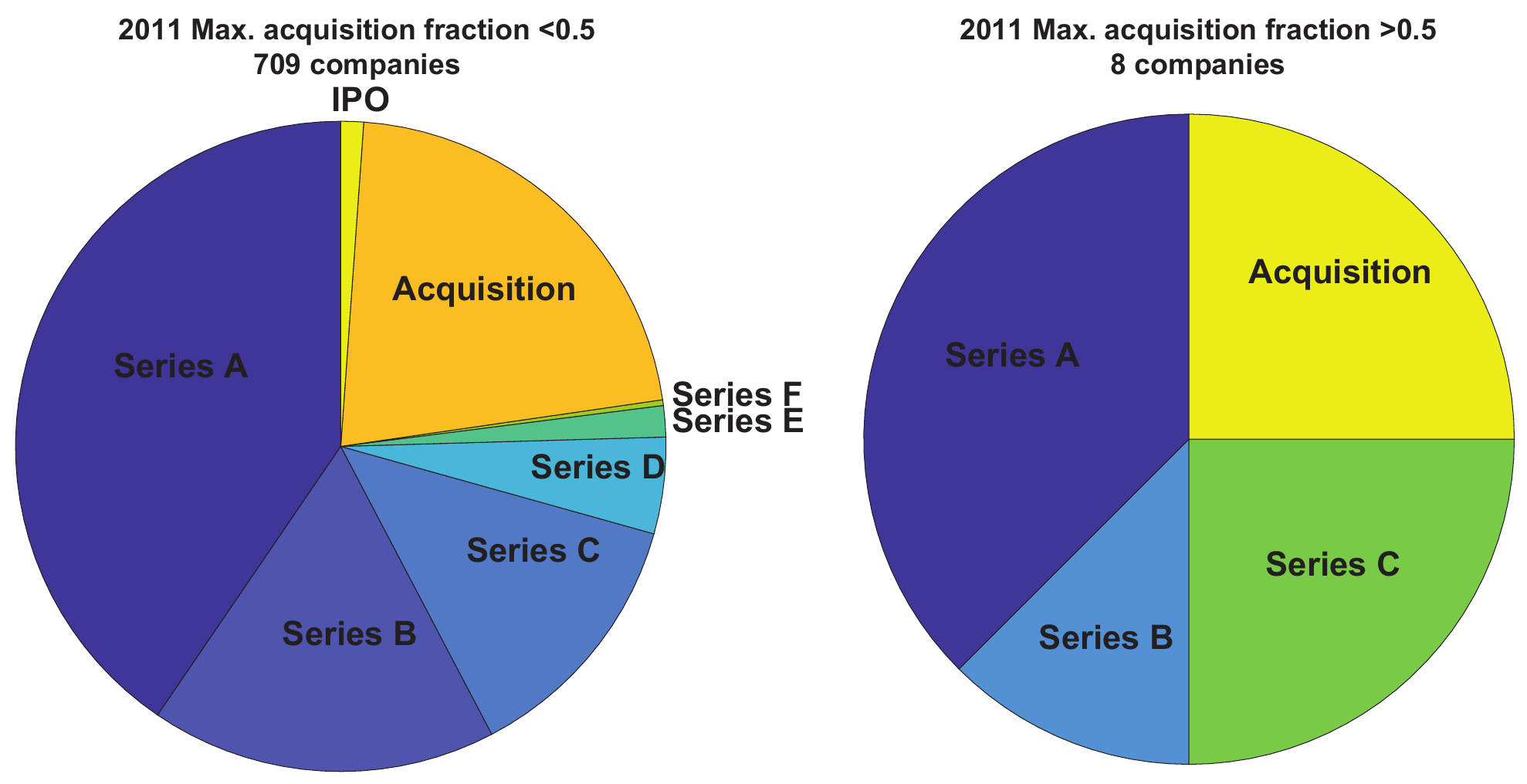}
	\end{center}
	\caption{Pie graphs of the funding round distribution of companies with the \emph{maximum acquisition fraction} feature less than 0.5 (left) and greater than 0.5 (right).}
	\label{fig:maxacqfracpie}
\end{figure}

To visualize the impact of the investor features, we plot the distribution of companies final funding round conditioned on the  \emph{maximum acquisition fraction} feature (rounded to the nearest whole number) in Figure \ref{fig:maxacqfracpie}.  We can see that the exit rate is slightly higher for companies with a high value for this feature.  To be precise, among companies where this feature is less than 0.5 (709 companies), 22\% are acquisitions and 1\% are IPOs, while among companies where this feature is greater than 0.5 (8 companies), 25\% are acquisitions.  This is a small difference, but still seems to agree with previous studies \citep{nanda2017persistent}.  However, because of the small number of companies with a high value for this feature,  one cannot make any strong conclusions from Figure \ref{fig:maxacqfracpie}.


\textbf{Competitors Data and Features.} We also construct a dynamic network of companies where edges between companies signify that they are competitors. Specifically, a directed edge $e_{ij}$ exists between company $i$ with baseline date $t_i$ and company $j$ with baseline date $t_j < t_i$ if at least one of company $i$ or $j$ listed the other as a competitor on Crunchbase or if both companies $i$ and $j$ listed the same company $k$ with baseline date $t_k < t_i$ as a customer on Crunchbase.

Using this network, we define features \emph{competitors A}, \emph{competitors B}, \emph{competitors C}, \emph{competitors D}, \emph{competitors E}, \emph{competitors F}, \emph{competitors acquisition}, and \emph{competitors IPO}. Each of these features for a company $i$ is calculated as the number of edges $e_{ij}$ pointing to companies $j$ whose highest round at time $t_i$ is the specified round for the feature divided by the out-degree of $i$. Additionally, for every company $i$ we include a feature called \emph{had competitor info} which is a binary indicator for whether the company self-reported it's competitors on Crunchbase. We include \emph{had competitor info} as a feature because we suspected that whether or not founders were willing to self-report their competitors is potentially a psychological factor which influences whether or not the founders are able to run a successful company.

\begin{figure}
	\begin{center}
    \includegraphics[scale=.35]{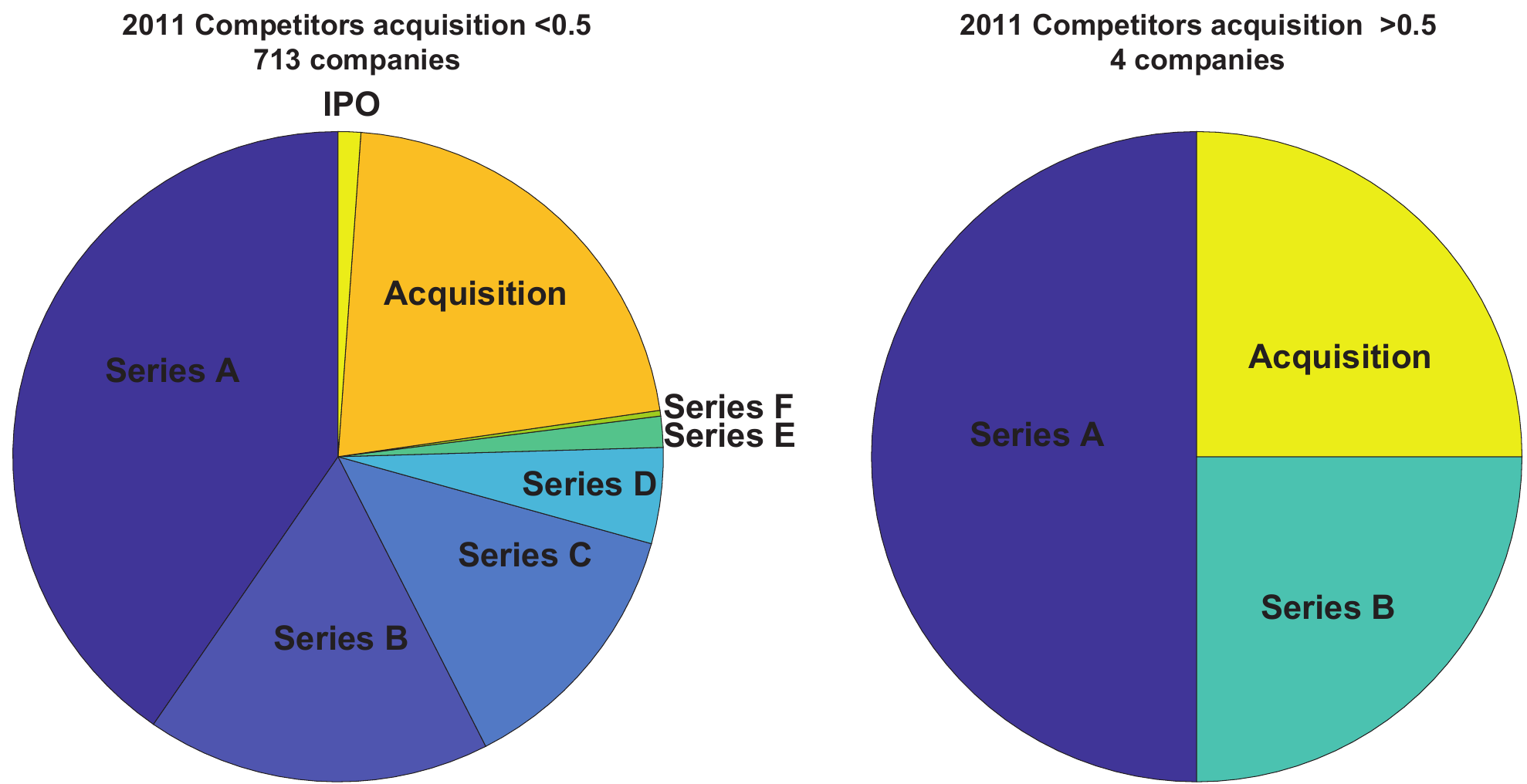}
	\end{center}
	\caption{Pie graphs of the funding round distribution of companies with the \emph{competitors acquisition} feature less than 0.5 (left) and greater than 0.5 (right).}
	\label{fig:compacqpie}
\end{figure}
To visualize the impact of the competitor features, we plot the distribution of companies final funding round conditioned on the  \emph{competitors acquisition} feature (rounded to the nearest whole number) in Figure \ref{fig:compacqpie}. We can see that the exit rate is slightly higher for companies with a high value for this feature.  Among companies where this feature is less than 0.5 (713 companies), 22\% are acquisitions and 1\% are IPOs, while among companies where this feature is greater than 0.5 (4 companies), 25\% are acquisitions.  As with the investment feature from Figure \ref{fig:maxacqfracpie}, the huge imbalance in companies prevents one from making any strong conclusions from Figure \ref{fig:compacqpie}.


\textbf{Leadership Data and Features.}
The leadership features of a company are derived from the Crunchbase and LinkedIn data for its founders and executives. We are particularly interested in measuring the experience, education, and ability of the company's leadership.

We first use the Crunchbase data to consider the employees, executives, and advisors of the company of interest. We construct indicator features \emph{job IPO}, \emph{job acquired}, \emph{executive IPO}, \emph{executive acquired}, \emph{advisory IPO}, and \emph{advisory acquired} which indicate if someone affiliated with the company has been part of a previous company that was either acquired or reached an IPO before our date of interest. In particular, for the \emph{job} variables we consider people who work for the company but are not executives or advisors, for the \emph{executive} variables we consider people who are labeled as an executive in the company, and for the \emph{advisor} variables we consider people who are labeled as advisors but not executives.

  Next we consider the features based on the LinkedIn data.  We took great care was taken to ensure that the information that we used from LinkedIn did not violate causality. In particular, we made sure that an investor could have known the information that we acquired from LinkedIn when they were considering investing in a startup.
  
   One LinkedIn feature that we use is \emph{previous founder} which
is the fraction of the leadership that had previously founded a company
before the given company's baseline date. 
We also use \emph{number of companies affiliated}, which is the average
number of companies each leadership member was affiliated with before
joining the given company.  A final pair of experience features is 
\emph{work overlap mean} and \emph{work overlap standard deviation}.
To construct these features, we calculate the Jaccard index of previous
companies for each pair of leadership members.  The Jaccard index is defined
as the intersection of previous companies divided by the union of previous companies
for the pair of members.  We then take the mean and standard deviation of these values
across all pairs of leadership members to obtain the two features. We chose to construct these feature using the LinkedIn data instead of the Crunchbase data, because we often found that the LinkedIn information suggested that a person had previous startup experience, when the Crunchbase data did not contain this information. We believe that this is due to the fact that Crunchbase is slightly biased to include companies that were somewhat successful. 

An education feature we use is \emph{from top school} which is the fraction
of leadership members that went to a top school.  This list of top schools
was created from a combination of known top rankings \citep{usnews} and our own
knowledge and is provided in Appendix \ref{app:schools}.  We also have features for the
highest education level of the leadership, measured by degree received.  
These features are \emph{high school},
\emph{bachelors}, \emph{master's}, and \emph{Ph.D.}  For each degree, we
measure the fraction of leadership members whose maximum education level
equals that degree to obtain the feature value. 

We also have features based on the education and academic major overlap.
These features are \emph{education overlap mean}, \emph{education standard deviation},
\emph{major overlap mean}, and \emph{major standard deviation}.  For each one
we calculate the relevant Jaccard index over all pairs of leadership members,
and then take the mean and standard deviation.  For education, the Jaccard index
is taken with respect to the schools attended by each member and for major, the Jaccard index
is taken with respect to the academic majors of each member.

A more complex feature we use is \emph{major company similarity} which captures
the similarity of the academic major of the leadership and the company sector.
We use the WordNet lexical database to create a semantic similarity score between each member's major
and the sector of their company \citep{wordnet}.  We use the Palmer-Wu similarity
score, which measures the similarity of words in a semantic network
based on the distances to their most common ancestor and to the root \citep{wu1994verbs}.
This score is zero for totally different words and one for equivalent words.
We average the Palmer-Wu major-sector similarity score over the leadership members to obtain the feature value.

A final feature we use is \emph{leadership age}.  This is simply
the average age of all leadership members when the company was founded.  
To estimate the age we assume that each member is 18 when finishing high-school
and 22 when finishing undergraduate education.
With this assumption, we set the age of a member to be equal to $22$ plus the company founding year minus the year the
member received/would have received an undergraduate degree. 

\begin{figure}
	\begin{center}
    \includegraphics[scale=.35]{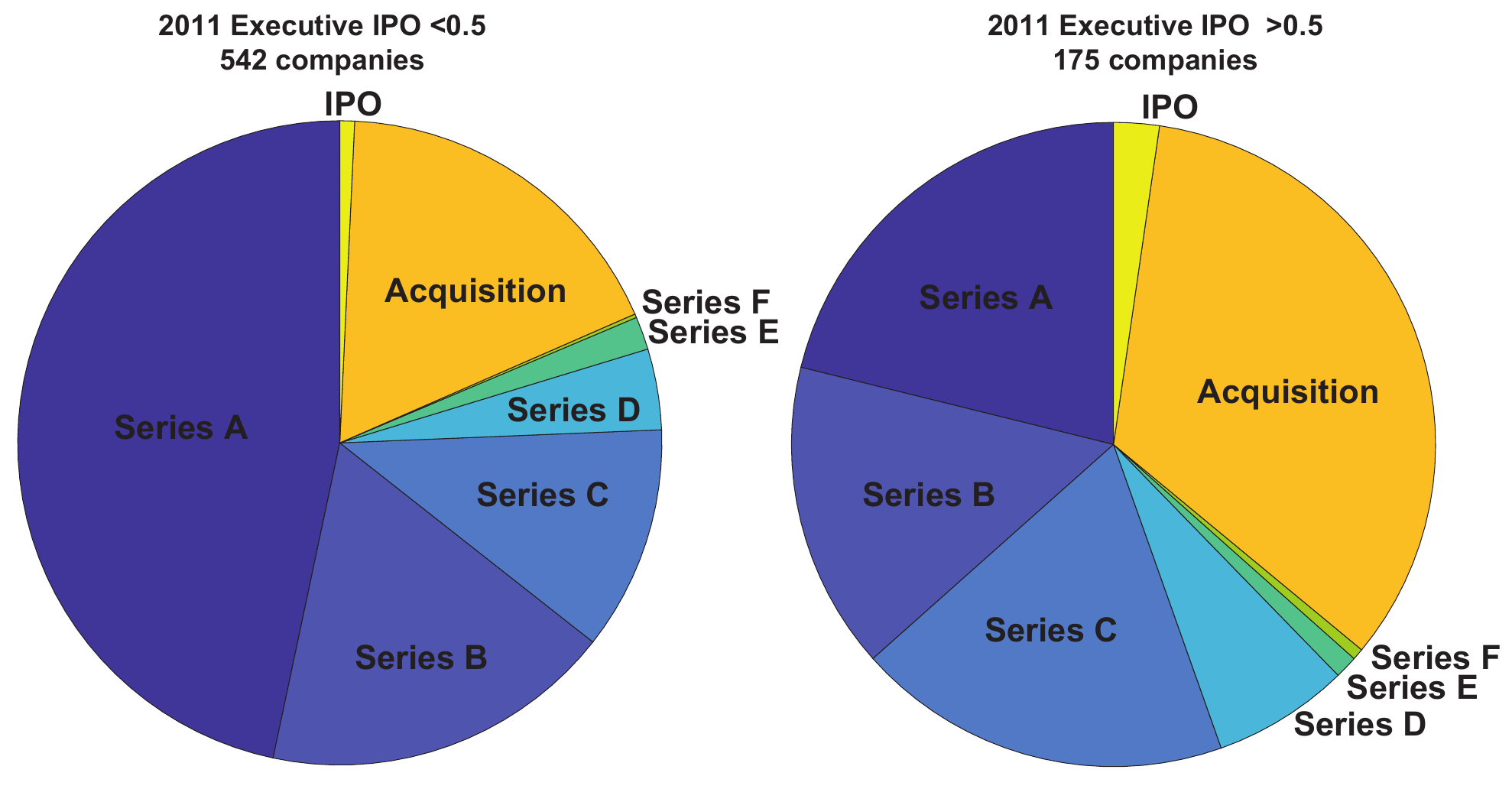}
	\end{center}
	\caption{Pie graphs of the funding round distribution of companies with the \emph{executive IPO} feature less than 0.5 (left) and greater than 0.5 (right).}
	\label{fig:execipoqpie}
\end{figure}

In order to understand the impact of the leadership features, we plot the distribution of companies by funding round with the investor features rounded to the nearest whole number. Shown in Figure \ref{fig:execipoqpie} is an example for the feature \emph{executive IPO}. From these graphs, we can see that there is a higher exit rate for companies with a high \emph{executive IPO} value.  The large difference in the exit rates suggests that previous success is a good indicator of future success.

\subsubsection{Imputation of Missing Data}
For many of the LinkedIn features we were unable to obtain data.  We still needed some value for these features to use the resulting feature vector in our models.
To do this we imputed the missing values.  We concatenated all the $M$ dimensional feature vectors
for $N$ companies, creating a $M \times N$ matrix with missing values.  To impute these missing
values, we use a matrix completion algorithm known as Soft-Impute which achieves the imputation
by performing a low-rank approximation to the feature matrix using nuclear norm regularization \citep{mazumder2010spectral}. 
Soft-Impute requires a regularization parameter and a convergence threshold.  To obtain the regularization parameter, we 
replace all missing values with zero and then calculate the singular values of the resulting
matrix.  We set the regularization parameter to be the maximum singular value of this filled in matrix divided by 100 and we set
the convergence threshold to 0.001.  We then apply Soft-Impute with these parameters to the incomplete
feature matrix to fill in the missing values.  This imputed feature matrix is used for all subsequent model fitting and prediction
tasks. Note that we repeat this process for each training set and testing set that we consider, to ensure that we are not violating causality.

\section{Model for Startup Company Funding}\label{sec:model}
 We now present a stochastic model for how a company reaches different funding rounds and the relevant probability calculations needed to construct portfolios.   Our model captures the temporal evolution of the funding rounds through the
use of a Brownian motion process. 


\subsection{Brownian Motion Model for Company Value}
We assume a company has a latent value process $X(t)$ which is
a Brownian motion with time-dependent drift $\mu(t)$ and diffusion coefficient $\sigma^2(t)$. The latent value process $X(t)$ is initiated with value $0$   when the company receives its first funding round.     The set of possible funding rounds that we use is $\mathcal R = \curly{\text{Seed},\text{A},\;\text{B},\;\text{C},\;\text{D},\;\text{E},\;\text{F},\;\text{Exit}}$.  We denote each funding round by an index $0\leq l \leq 7$, with $l=0$ corresponding to seed  funding, $l=1$ corresponding to series A funding, etc.  The final level is $l=7$, which corresponds to exiting.
  For each funding round $l$ we have a level $h_l\geq 0$, and the levels are ordered such that $h_{l-1}\leq h_l$.  We choose the spacing of these levels to be linear, but one is free to choose them arbitrarily.  In our model, we let $h_l = \Delta l$ for some $\Delta>0$.  
A company  receives round $l$ of funding when $X(t)$ hits $h_l$ for the first time, and we denote this time as $t_l$.
Therefore, the time to receive a new round of funding is the first passage time of a Brownian motion.

The first passage time distribution for a Brownian motion with arbitrary time-varying drift and diffusion terms is difficult to solve. In particular, one must solve the Fokker-Planck equation with an appropriate boundary condition \citep{bm}. However, we can solve the Fokker-Planck equation exactly with the appropriate boundary condition when the ratio of the drift to the diffusion is constant in time.
Therefore, we assume that the drift and diffusion terms are of the form $\mu(t) = \mu_0 f(t)$ and $\sigma^2(t) = \sigma^2_0 f(t)$, where $\mu_0$, $\sigma^2_0$, and $f(t)$ are appropriately chosen. Under these assumptions, we have the following standard result for the first passage time distribution.

\begin{thm}\label{thm:FP} [\cite{bm}]
For a Brownian motion $X(t)$ with drift $\mu(t) = \mu_0 f(t)$, diffusion $\sigma^2(t) = \sigma^2_0 f(t)$, and initial value $X(v_0) = 0$,  let $V_\alpha = \inf_{t>v_0}\curly{X(t)\geq \alpha}$ denote the first passage time to level $\alpha > 0$ after time $v_0$. Then, the probability density function (PDF) of $V_\alpha$ is 
\begin{align}
\label{eq:f0pdf}
f_0(v;v_0, \mu(t),\sigma(t),\alpha) & = \frac{\sigma^2(v) \alpha }{ \sqrt{16 \pi {S}^3}} e^{ - \tfrac{(\alpha - M)^2}{4S}}
\end{align}
and the cumulative distribution function (CDF) is 
\begin{align}
F_0(v;v_0, \mu(t),\sigma(t),\alpha) & = \Phi\paranth{\frac{M- \alpha}{\sqrt{2S}}}
+\exp\curly{\frac{M\alpha}{S}}
\Phi\paranth{\frac{-(M + \alpha)}{\sqrt{2S}}}\label{eq:cdf}.
\end{align}
where $M = \int\limits_{v_0}^v \mu(s) \text{d}s$, $S = \frac{1}{2} \int\limits_{v_0}^v \sigma^2(s) \text{d}s$,
and $\Phi(\cdot)$ is the standard normal CDF.  
\end{thm}
%




\subsection{Modeling Drift and Diffusion}
\label{sec:driftdiffusionmodeling}

From Section \ref{sec:data}, recall that companies which succeed typically take a constant amount of time to reach each successive funding round, which motivates these companies having a positive and constant drift coefficient for some time period. Additionally, many companies succeed in reaching early funding rounds somewhat quickly, but then they fail to ever reach an exit. This motivates these companies having a drift that decreases in time. Lastly, as time gets very large, a company that has not achieved an exit will most likely not achieve a new funding round. This translates to our latent value process moving very little as time gets large, which can be modeled by having the drift and diffusion terms moving toward zero. To incorporate these properties, we use the following model for the drift and diffusion coefficients for a company $i$:
\begin{align}
\mu_i(t) &= \mu_{i0} \left( \mathbf{1}\left\{  t \leq \nu   \right\} + e^{ - \tfrac{t-\nu}{\tau}} \mathbf{1}\left\{  t > \nu \right\}  \right)\label{eq:mu_time} \\ 
\sigma_{i0} ^2(t) &= \sigma_{i0}^2 \left( \mathbf{1}\left\{  t \leq \nu   \right\} + e^{ - \tfrac{t-\nu}{\tau}} \mathbf{1}\left\{  t > \nu \right\}  \right)\label{eq:sigma_time}
\end{align}
where $\mu_{i0} $, $\sigma_{i0} ^2$, $\nu$, and $\tau$ are appropriately chosen based on the data.  Under this model,
the drift and diffusion are constant for a time $\nu$, after which they decay exponentially fast with time
constant $\tau$.

In our model, every company will have the same $\nu$ and $\tau$. However, each company will have a different drift term $\mu_{i0} $ and diffusion term $_{i0} $ that will be determined by its features. For a company $i$ we define a feature vector $\mathbf x_i \in \mathbbm R^M$, a drift $\mu_{i0}$,
and a diffusion $\sigma_{i0}^2$. We define a parameter vector $\beta_y \in \mathbb R^M$ for year $y$. A company $i$ that has baseline year (earliest funding year) $y$ has $\mu_{i0} = \beta_y^T \mathbf x_i$. Furthermore, we set $\beta_{y+1} = \beta_y + \epsilon$ where $\epsilon = \left[ \epsilon_1, \ldots, \epsilon_M \right]$ and $\epsilon_i$ is normally distributed with zero mean and variance $\delta_i^2$ for $1 \leq i \leq M$. This time-varying coefficient model allows us to capture any dynamics in the environment which could increase or decrease the importance of a feature. For instance, consider a feature which is whether or not the company is in a certain sector. The coefficient for this feature would change over time if the market size for this sector changes. Additionally, this time-varying model allows us to capture uncertainty in future values of the drift weights, which we will utilize in Section \ref{sec:exit_prob} to construct the joint distribution of first passage times for the companies.

Additionally, we found that  funding rounds data varies substantially between companies. For instance, some companies will frequently achieve early funding rounds very quickly, but then fail to ever reach an exit. To model this phenomenon, we introduce heteroskedasticity into the model and allow each company to have its own diffusion coefficient.  We define another parameter vector $\gamma \in \mathbb R^M$, and we let $\sigma_{i0}^2 = {g\left( \gamma^T \mathbf{x}_i \right)}^2$.  The diffusion coefficient must be non-negative.  Additionally, we would like it to approach zero as $\gamma^T \mathbf{x}_i$ becomes more negative, while for postive values, we would like it to approach $\gamma^T \mathbf{x}_i$.  To do this, 
  we let $g(z)$ be the positive region of the hyperbola with asymptotes at $z = 0$ and $g(z) = z$.  Precisely, 
\begin{align*}
g(z) & = \frac{1}{2}\paranth{z+\sqrt{z^2+\frac{1}{4}}}.
\end{align*}
This function is zero for large negative values of $z$ and approaches $z$
for large positive values of $z$, as we require.  

We considered many other models for the drift and diffusion terms. We found that the model presented here performed well with respect to fitting the data, and additionally we found this model to be intuitive and interpretable. Therefore, we will restrict our future results to those that were obtained using this particular model.


\subsection{Data Likelihood}
\label{sec:datalikelihood}

Using our definitions of drift and diffusion, we can now define the likelihood for the observed data of a given company. For notation, suppose that we observe a company at time $t_{obs}$.  Before this time it reaches a set of funding rounds with indices $\mathbf i = \curly{i_1,i_2,...i_L}$ . The random variables corresponding to the
times of these rounds are $\mathbf t=\curly{t_{i_1},t_{i_2},...,t_{i_L}}$ (we assume $i_0=0$ and normalize the times so $t_{i_0}=0$).  There may be missing data, so the funding rounds
are not necessarily consecutive. For a given company $c$, we use $\mathbf t_c$ for the times of the funding rounds, $\mathbf x_c$ for the feature vector, and $c_y$ for the baseline year.

If an exit is observed for a company $c$, then $i_L = 7$ and by the independent increments property and translation invariance of Brownian motion, the likelihood is given by
\begin{equation}\label{eq:exitlikelihood}
\mathbb{P}(\textbf{t}_c, t_{obs}, \textbf{x}_c, h_{i_L} = 7\Delta | \beta_{c_y}, \gamma, \delta, \nu, \tau) = \prod\limits_{l=1}^L f_0 (t_{i_l} ; t_{i_{l-1}} , \mu_c(t), \sigma_c(t), h_{i_l} - h_{i_{l-1}}),
\end{equation}
where $f_0$ is defined in equation \eqref{eq:f0pdf} and $\mu_c(t)$ and $\sigma_c(t)$ are given by equations \eqref{eq:mu_time} and \eqref{eq:sigma_time}, respectively. 

If the company does not exit before $t_{obs}$, then $i_L < 7$ and there is a censored inter-round time which is greater than or equal to $t_{obs}-t_{i_L}$.   In this case the likelihood is given by
\begin{multline}\label{eq:nonexitlikelihood}
\mathbb{P}(\textbf{t}_c, t_{obs}, \textbf{x}_c, h_{i_L} < 7\Delta | \beta_{c_y}, \gamma, \delta, \nu, \tau) =  \\ \left( 1 - F_0(t_{obs}; t_{i_L}, \mu(t), \sigma(t), \Delta )   \right)\prod\limits_{l=1}^L f_0 (t_{i_l}; t_{i_{l-1}}, \mu_c(t), \sigma_c(t), h_{i_l} - h_{i_{l-1}}),
\end{multline}
where $f_0$ and $F_0$ are defined in equations \eqref{eq:f0pdf} and \eqref{eq:cdf} respectively.

For simplicity in notation, let the set of companies be $C$ and let $\textbf{T} = \{\textbf{t}_c, \textbf{i}_c, \{h_{i_j} | \forall i_j \in \textbf{i}_c \} | \forall c\in C$\} be the set of time series information (funding rounds levels and time to reach each level) for all companies $c$ in our data. Furthermore, let $\textbf{X}$ be the data matrix containing the features for all companies in our data.

We use the following additional definitions:   $C_e$ is the set of companies in our data that have exited, $C_f$ is the set of companies in our data that have not exited, and $\beta$ is the set of all $\beta_y$.  Using this notation we can write the full data likelihood as
\begin{multline}
\label{eq:fulldatalikelihood}
\mathbb{P}(\textbf{T}, \textbf{X} |  \beta, \gamma, \delta, \nu, \tau) = \\ \Big(\prod_{c \in C_e}\mathbb{P}(\textbf{t}_c, t_{obs}, \textbf{x}_c, h_{i_L} = 7\Delta | \beta_{c_y}, \gamma, \delta, \nu, \tau)\Big) \Big(\prod_{c \in C_f}\mathbb{P}(\textbf{t}_c, t_{obs}, \textbf{x}_c, h_{i_L} < 7\Delta | \beta_{c_y}, \gamma, \delta, \nu, \tau)\Big).
\end{multline}


\section{Bayesian Model Specification}
\label{sec:bayesmodelsection}

In this section we present four variations of the Brownian motion model using a Bayesian  approach which we focus our analysis on for the remainder of the paper. Throughout this section, we use the subscript $i$ to denote the $i$th component of a parameter vector, and denote $\mathcal N(\mu,\sigma^2)$ as a normal distribution with mean $\mu$ and variance $\sigma^2$, $\mathcal IG(a,b)$ as an inverse gamma distribution with shape parameter $a$ and scale parameter $b$, ${Exp}(\lambda)$ as an exponential distribution with mean $\lambda^{-1}$, and $\mathcal U(a,b)$ as a uniform distribution on $\bracket{a,b}$.  Prior distributions are generally chosen to be uninformative and conjugate when possible.  

\subsection{Homoskedastic Model}\label{sec:homoskedasticmodel}
The first variation of the model that we consider is a homoskedastic model. For this variation, we let $\sigma_0^2$ be the same for all companies rather than be a function of the company features.  This is done by having the parameter vector $\gamma$ be zero except for the constant feature. 
The posterior distribution of the parameters is as follows:
\begin{equation}
\label{eq:homokedasticposterior}
\mathbb{P}(\beta, \sigma_0^2, \delta, \nu, \tau | \textbf{T}, \textbf{X}) \propto \mathbb{P}(\textbf{T}, \textbf{X} |  \beta, \sigma_0^2, \delta, \nu, \tau) \Big(\prod_{y=1}^Y\mathbb{P}(\beta_{y} | \beta_{y - 1})\Big)\mathbb{P}(\beta_0)\mathbb{P}(\sigma_0^2)\mathbb{P}(\delta)\mathbb{P}(\nu)\mathbb{P}(\tau), 
\end{equation}
with prior distributions defined as follows:
\begin{equation}
\label{eq:homoskedasticpriors}
\begin{split}
\beta_{yi} \sim \mathcal{N}(\beta_{(y - 1)i}, \delta_i)\\
\beta_{0i} \sim \mathcal{N}(\mu_\beta, \sigma^2_\beta)\\
\sigma_0^2 \sim \mathcal{IG}(a_{\sigma_0^2}, b_{\sigma_0^2})\\
\delta_i \sim Exp(\lambda_0)\\
\nu \sim \mathcal{U}(a_\nu, b_\nu)\\
\log\tau \sim \mathcal{U}(a_\tau, b_\tau).
\end{split}
\end{equation}
The complete specification of the priors and hyperpriors can be found in Appendix \ref{app:samplingdetails}.

\subsection{Heteroskedastic Model}\label{sec:heteromodel}

The second model we introduce is a heteroskedastic model with drift and diffusion as described in Section \ref{sec:driftdiffusionmodeling}.
For this model, the posterior distribution of the parameters given the data over $Y$ total years can be written as:
\begin{equation}
\label{eq:heteroskedasticposterior}
\mathbb{P}(\beta, \gamma, \delta, \nu, \tau | \textbf{T}, \textbf{X}) \propto \mathbb{P}(\textbf{T}, \textbf{X} |  \beta, \gamma, \delta, \nu, \tau) \Big(\prod_{y=1}^Y\mathbb{P}(\beta_{y} | \beta_{y - 1})\Big)\mathbb{P}(\beta_0)\mathbb{P}(\gamma)\mathbb{P}(\delta)\mathbb{P}(\nu)\mathbb{P}(\tau),
\end{equation}
with prior distributions defined as follows:
\begin{equation}
\label{eq:heteroskedasticpriors}
\begin{split}
\beta_{yi} \sim \mathcal{N}(\beta_{(y - 1)i}, \delta_i)\\
\beta_{0i} \sim \mathcal{N}(\mu_\beta, \sigma^2_\beta)\\
\gamma_i \sim \mathcal{N}(\mu_\gamma, \sigma^2_\gamma)\\
\delta_i \sim Exp(\lambda_0)\\
\nu \sim \mathcal{U}(a_\nu, b_\nu)\\
\log\tau \sim \mathcal{U}(a_\tau, b_\tau).
\end{split}
\end{equation}
The specification of the priors can be found in Appendix \ref{app:samplingdetails}.

\subsection{Robust Heteroskedastic Model}
\label{sec:robustheteromodel}
One issue  with our data is that there is a significant class imbalance, with only a few percent
of the companies having exits.  This can lead to model estimation results which have poor performance in metrics
related to true positive and false positive rates.  For instance, a good model fit could be one which does
not predict that any company exits.  This would have a low false positive rate, but a near zero true positive rate.
To overcome this challenge, one can use techniques such as oversampling.  This modifies the training data by resampling
extra copies of the minority class, which in our case are companies that exit.  The drawback of this approach is that
it can lead to poor out of sample performance because the model will overfit on the resampled data points.  One popular technique that overcomes this issue is
known as synthetic minority over-sampling technique (SMOTE) which adds noise to the feature vector of the resampled
data points \citep{chawla2002smote} .  The inclusion of noise prevents overfitting from occurring.  

For our data, we use an oversampling technique which while not equivalent, is similar in spirit to SMOTE.
We will see in Section \ref{sec:portfoliosection} that our technique does result in better performance in terms of portfolio
exit rates, which can be viewed as true positive rates.  Our oversampling technique is as follows.
In the training data we resample each exit once.  However, we do not keep all the times of the different funding rounds.
Instead, the only information we keep regarding the funding rounds is the fact that an exit occurred before the observation
time, but we do not specify when it occurred.  For a company $c$, this results in a likelihood function equal to $ F_0(t_{obs}; t_{i_0}, \mu_c(t), \sigma_c(t), 7\Delta)$
for each resampled company, which is simply the probability the company exited sometime between its initial date $t_{i_0}$ and the observation time $t_{obs}$.  Since we resample each exiting company once, another way to write the likelihood for this model is to assume there was no resampling, but instead modify the likelihood of the exiting companies to
\begin{equation}
\label{eq:likelihoodexitrobust}
\mathbb{P}(\textbf{t}_c, t_{obs}, \textbf{x}_c, h_{i_L} = 7\Delta | \beta_y, \gamma, \delta, \nu, \tau) =  F_0(t_{obs}; t_{i_0}, \mu_c(t), \sigma_c(t), 7\Delta)\prod_{l=1}^Lf_0(t_{i_l}; t_{i_{l - 1}}, \mu_c(t), \sigma_c(t), h_{i_l} - h_{i_{l - 1}}).
\end{equation}
Here we have just multiplied the likelihood of the funding round times by the probability of an exit.  This likelihood is equivalent
to resampling the exits and ignoring their funding round times.   The model parameters of this model are the same as the heteroskedastic model, so we use the same posterior distributions described in Section \ref{sec:heteromodel}.  

We have found that this model performs substantially better than the unmodified heteroskedastic model on the portfolio selection problem as shown in Section \ref{sec:portfoliosection}. We will refer to this model as the robust heteroskedastic model while we refer to the unmodified model as the heteroskedastic model for the remainder of the paper.

\subsection{Model Estimation}
\label{sec:estimation}

To estimate the models we first select a value for the level spacing $\Delta$ and for the observation year $t_{obs}$.  We choose $\Delta=10$ and  we set $t_{obs}$ equal to
December 31st of the chosen year.  We include in the training set
all companies founded between the year 2000 and $t_{obs}$.  All data used
in the model estimation must have been available before $t_{obs}$, otherwise
we do not include it.  For instance, the observed
funding round times occur before $t_{obs}$.  If a company had an exit
after $t_{obs}$, this information is not included during model estimation.
  Also, all company features are constructed using data available when the company received its first funding round.  We do this because these features are what
 would have been available to someone
who wanted to invest in the company in an early round, which
is the investment time frame on which we focus.


We used blocked Gibbs sampling to estimate the model with each parameter vector ($\beta_y$, $\gamma$, $\delta$) in it's own block and all scalar values such as $\nu$, and $\tau$ in their own block. We found blocked Gibbs sampling to have the best trade-off between model estimation quality and speed. The exact details for how we sampled each parameter block are provided in Appendix \ref{app:samplingdetails}.

In our estimation, we ran five chains each of which contained 25,000 samples. We used a burn-in period of 15,000 samples and took every tenth sample beyond that to reduce autocorrelation in the samples. Convergence was assessed using the Gelman-Rubin criterion \citep{gelmanrubin}.

\section{Parameter Estimation Results}
\label{sec:estimationresults}



Here we present the results of estimating the different models (referred to as homoskedastic, heteroskedastic, and robust heteroskedastic) presented in Section \ref{sec:bayesmodelsection}. First, we present model fit scores for each model. Then, we analyze the statistical significance of the estimated parameter values in order to understand which features are most important in determining the exit probability of a company. In Section \ref{sec:portfoliosection}, we will show that the robust heteroskedastic model gives the best startup company portfolios. Therefore, among the homoskedastic, heteroskedastic, and robust heteroskedastic models, we show parameter estimation results for the robust heteroskedastic model in this section. In all sections describing parameter value estimation, we focus our analysis on the model trained with observation year 2010.

\subsection{Model Fit Scores}
\label{sec:modeldic}

We calculated the deviance information criterion (DIC) \citep{ref:dic} to evaluate the fit of each model.  The DIC measures fit using a negative log-likelihood on the data, but penalizes models with too many degrees of freedom (parameters).  A smaller DIC score indicates a better model fit. We show the DIC scores for the models in Table \ref{table:dicvals}.  Among the  models, the heteroskedastic model had the lowest DIC, and thus the best tradeoff between data fit and number of parameters.  It does better than the homoskedastic model, which only allowed companies
to differ in the mean of the Brownian motion. This shows that using company specific diffusion coefficients improves model fit.  

\begin{table}[h]
	\centering
	\caption{The deviance information criterion scores for each  model trained with observation year 2010.}
	\label{table:dicvals}
	\begin{tabular}{|c|c|}
		\hline
		\textbf{Model Type} & \textbf{DIC Score}\\	\hline
			Homoskedastic & 41,026.51 \\ \hline
			Heteroskedastic & 38,650.72\\ \hline
			Robust Heteroskedastic & 52,741.08 \\ \hline
	\end{tabular}
\end{table}
The robust heteroskedastic model had the highest DIC score.  This may be due to the fact that when we include the extra terms for the exit probabilities, it is equivalent to adding extra data points.   That can result in a lower likelihood value since the probabilities are less than one. This means we cannot directly compare this model with the other models based on the DIC.   However, despite having a higher DIC, we will see in Section \ref{sec:portfoliosection} that this model actually does the best in terms of picking  companies for portfolios.  The out of sample portfolio performance is a better way to compare the different models than the DIC.




\subsection{Robust Heteroskedastic Model}
We now study the robust heteroskedastic model estimation results in more detail, as this is the best performing model in terms of portfolio construction (see Section \ref{sec:portfoliosection}). We present the results using boxplots of the posterior samples for the parameters.  We only include features which do not contain zero in their 90\% posterior credibility interval, and hence show significance in the Bayesian sense.

\begin{figure}
	\centering		\includegraphics[scale=0.40]{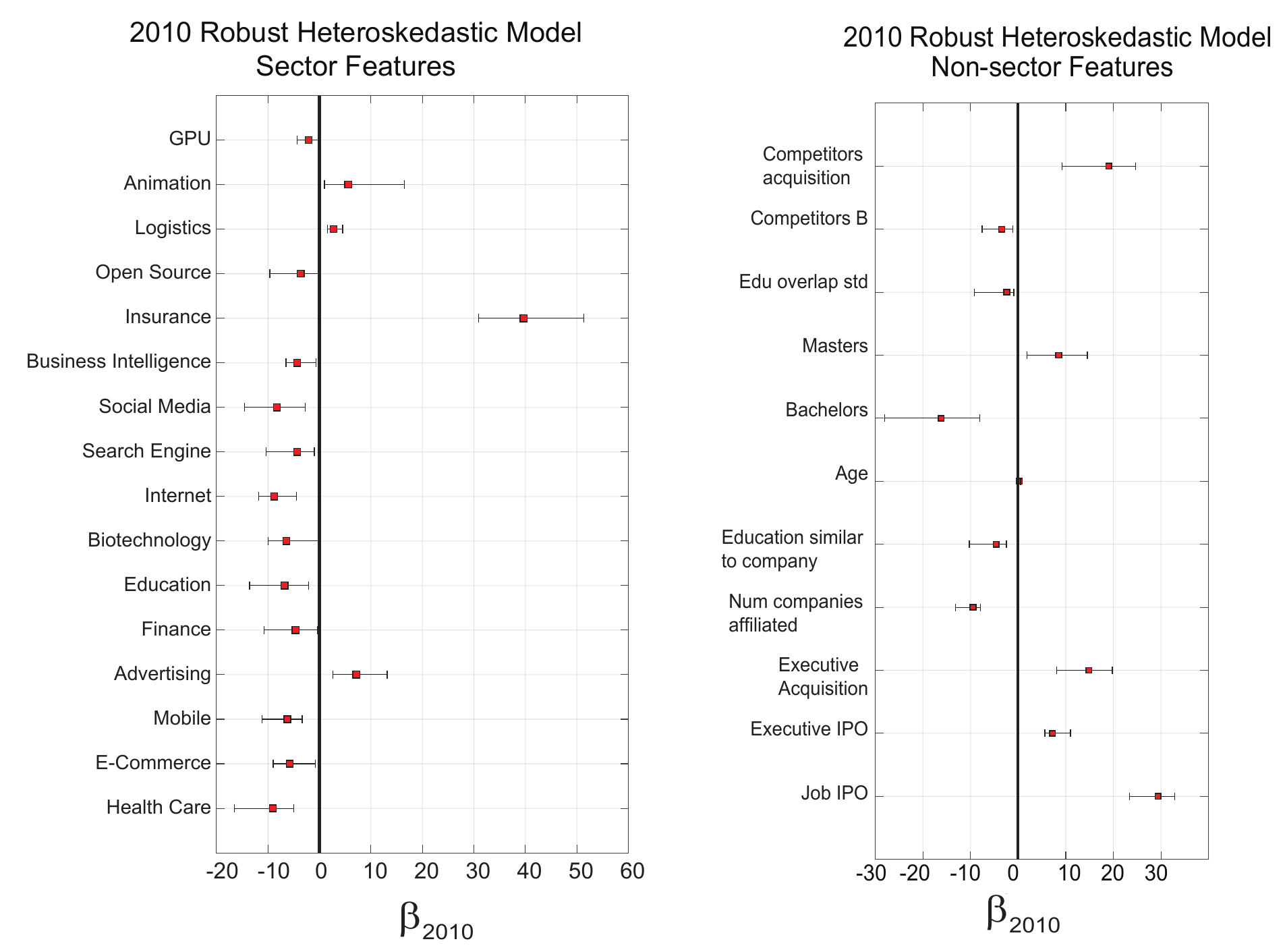}

	\caption{Boxplots of the $\beta_{2010}$ posterior sample values for significant sector (left) and non-sector (right) features.  Significance is defined as not having zero in the 90\% posterior credibility interval.}
	\label{fig:m3betasbox}
\end{figure}

  Shown in Figure \ref{fig:m3betasbox} are boxplots of the $\beta_{2010}$ posterior samples, with  separate plots for sector non-sector features.  We see that industries which typically are not cyclical and known for being constant sources of revenue such as \emph{animation}, \emph{insurance}, and \emph{advertising} are positive and statistically significant, and hence associated with a higher company drift. However, sectors which are known for being highly competitive or over-regulated such as \emph{social media}, \emph{biotechnology}, \emph{finance}, and \emph{healthcare} are negative and statistically significant, and hence associated with a lower company drift. From the  boxplot which shows non-sector features, we see that having previous entrepreneurship success among the executives of a company yields a higher company drift (the \emph{executive IPO} and \emph{executive acquisition} features).

\begin{figure}
	\centering
	\includegraphics[scale=0.40]{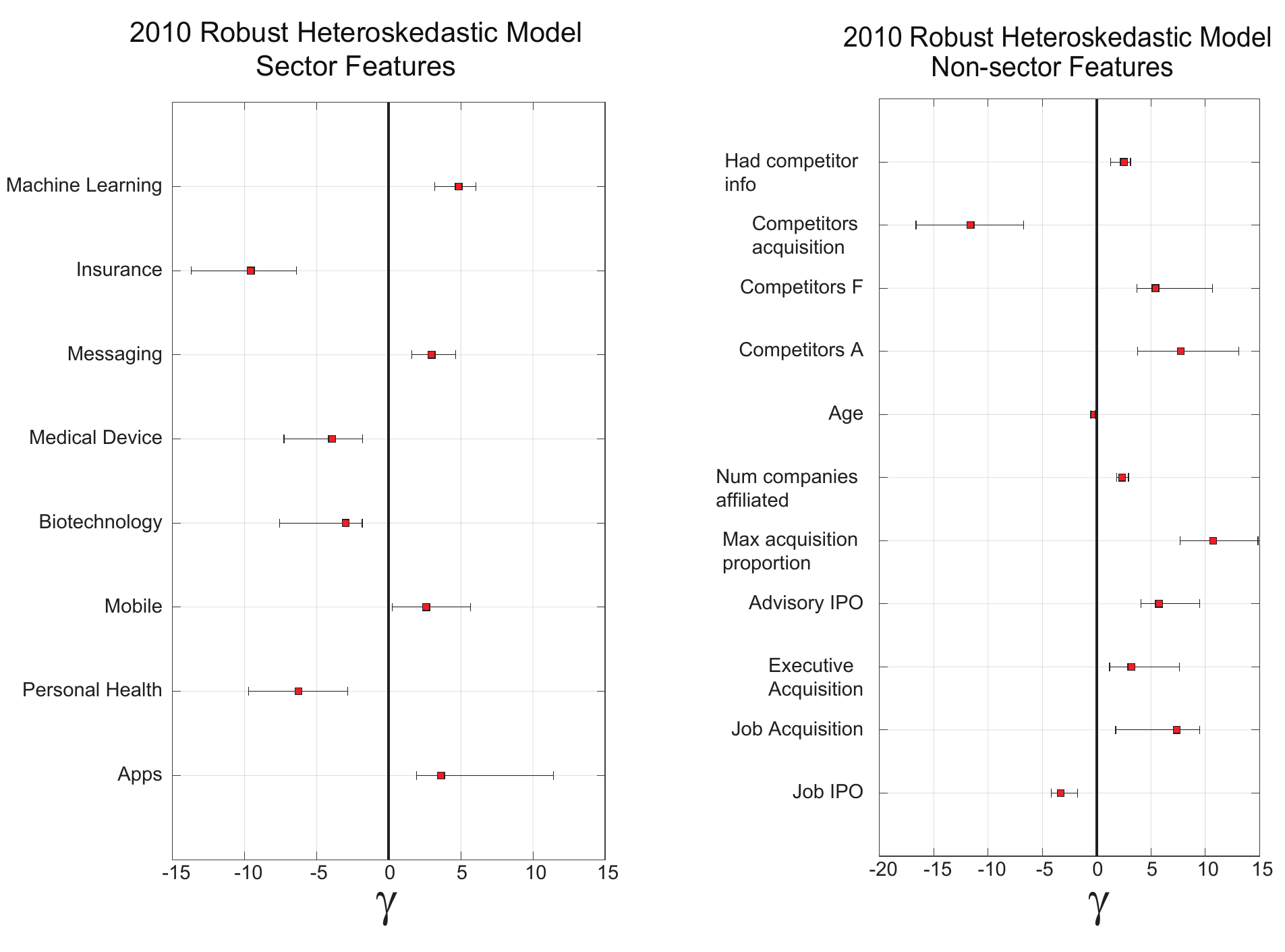}
	\caption{Boxplots of the $\gamma$ sample values for significant sector (left) and non-sector (right) features.  Significance is defined as not having zero in the 90\% posterior credibility interval.}
	\label{fig:m3gammasbox}
\end{figure}

Shown in Figure \ref{fig:m3gammasbox} are boxplots of the $\gamma$ posterior samples. We do not see any clear trends for the
sector features for the the diffusion.  However, for the non-sector features we see that previous success increases
the diffusion coefficient, similar to what was seen with the $\beta_{2010}$.  The main conclusion we can draw from this analysis is that previous success increases drift and diffusion, and hence the probability of an exit.  This conclusion
aligns with common knowledge in the venture capital industry, but our model provides a way to quantify this intuition.

Recall that for the time-varying $\beta$ parameters we have the standard deviation
$\delta$ which characterizes the variation of these parameters over time.
We show the top coefficient of variation, defined as the posterior median delta $\delta$ divided by
the absolute value of the average of the posterior medians of the $\beta$ samples over all years, for both the sector and non-sector features in Tables \ref{table:m3cvar_sector} and \ref{table:m3cvar_nonsector}.
The mean values of $\beta$ for the parameters with the largest coefficients of variation are quite small. Therefore, the coefficients that show large amounts of temporal volatility likely do not strongly impact the drift and exit probability of a company.

Finally, we show the statistics of the the timing parameters in Table \ref{table:timing}. We find that the posterior distribution of $\nu$ has a mean of 3.21 years with fifth and ninety-fifth quantiles at 2.99 years and 4.00 years respectively. We find that $\tau$ has a posterior mean of 3.50 years with fifth and ninety-fifth quantiles at 2.71 years and 4.03 years respectively. The estimated value of $\nu$ makes sense because the majority of exits in our data are acquisitions and as we saw from our data analysis, acquisitions take approximately three to four years to happen. Beyond the four year mark, we see decay in both the drift and diffusion of the company. The estimated value of $\tau$, the rate at which the drift and diffusion decay also makes sense based on our data because after seven years, very few companies exit.

\begin{table}
\centering
\caption{Posterior statistics of $\beta$, $\delta$, and the coefficient of variation for the sector features ranked by decreasing magnitude of the coefficient of variation.}
\label{table:m3cvar_sector}
\begin{tabular}{|c|c|c|c|}
\hline
\textbf{Sector feature} & \textbf{Avg. Median }$\beta$& \textbf{Median }$\delta$ & \textbf{Coefficient of variation}\\ \hline
Facebook& -0.0125 & 1.558 & 124.51\\ \hline
Computer& -0.0517 & 3.911 & 75.71\\ \hline
Open source& -0.0628 & 2.1379 & 34.05\\ \hline
Biotechnology& -0.1699 & 3.0617 & 18.02\\ \hline
Health diagnostics& 0.1302 & 1.4768 & 11.34\\ \hline
\end{tabular}
\end{table}

\begin{table}
\centering
\caption{Posterior statistics of $\beta$, $\delta$, and coefficient of variation for the non-sector features features ranked by decreasing magnitude of the coefficient of variation.}
\label{table:m3cvar_nonsector}
\begin{tabular}{|c|c|c|c|}
\hline
\textbf{Non-sector feature} & \textbf{Avg. Median }$\beta$& \textbf{Median }$\delta$ & \textbf{Coefficient of variation}\\ \hline
High school & -0.2795 & 1.951 & 6.979\\ \hline 
Max ipo proportion & 0.2777 & 1.735 & 6.246\\ \hline 
Competitors D & 0.2988 & 0.042 & 4.599\\ \hline 
Education overlap standard deviation & 0.6290 & 2.580 & 4.102\\ \hline 
Phd & -0.7373&  2.850 & 3.865\\ \hline 	
\end{tabular}
\end{table}

\begin{table}
\centering
\caption{Statistics of the posterior distribution of $\nu$ and $\tau$. The posterior median is shown with the fifth and ninety-fifth quantiles in parentheses.}
\label{table:timing}
\begin{tabular}{|c|c|}
\hline
Parameter & Median (5th quantile, 95th quantile)\\ \hline 
\textbf{$\nu$} & 3.21 (2.99, 4.00)\\ \hline 
\textbf{$\tau$}& 3.50 (2.71, 4.03)\\ \hline
\end{tabular}
\end{table}

\section{Picking Winners Portfolio}\label{sec:problem}
We now present the picking winners portfolio framework that we will use to evaluate
the performance of our models.  We describe
the framework in general and provide theoretical analysis of the portfolio construction algorithm.  We also compare our framework to the well known log-optimal portfolio framework.

\subsection{Objective Function}
Consider a set of events $\mathcal E=\curly{E_i}_{i=1}^m$ defined on a probability space $(\Omega,\mathcal F,\mathbf P)$, with sample space $\Omega$, $\sigma$-algebra $\mathcal F$, and probability measure $\mathbf P$.  Each event $E_i$ corresponds to item $i$ winning. The aim is to  select a subset $\mathcal S\subseteq \mathcal [m] = \left\{ 1, \ldots, m  \right\}$ of size $k$ such that we maximize the probability that at least one item in $\mathcal S$ wins, which is given by $\mathbf P\paranth{\bigcup_{i\in \mathcal S}E_i}$.  For notational convenience, we denote $U(\mathcal S) = \mathbf P\paranth{\bigcup_{i\in \mathcal S}E_i}$.  The optimization problem we want to solve is then
 \begin{align}
 	\max_{\mathcal S\subseteq \mathcal [m], |\mathcal S|=k}~U(\mathcal S). \quad 
 \end{align}
In general, solving this optimization problem can be difficult.  In fact, we have the following result.
\begin{thm}\label{thm:nphard}
Maximizing $U(\mathcal S)$ is NP-hard.
\end{thm}
Despite the computational complexity in maximizing $U(\mathcal S)$, the objective function
possesses some nice properties which we can use to obtain good solutions efficiently, which
are given by the following well-known result.
\begin{lem}\label{thm:submodular}
The function $U(\mathcal S)$ is non-negative, non-decreasing and submodular.
\end{lem}
Monotonically non-decreasing submodular functions have an important property which is that one can bound the sub-optimality of a greedy solution. Such a solution is obtained by sequentially adding elements to the set $\mathcal S$ such that each new addition results in the largest marginal increase in the objective function.  Formally, let $\mathcal S_G =\paranth{f_1,f_2,...,f_k}$ be the greedy solution of the optimization problem. The greedy solution is an ordered set, with item $f_i$ being added on the $i$th step of the greedy optimization.  Let $S_G^i =\paranth{f_1,f_2,...,f_i}$ be the solution obtained after the $i$th item is added to $S_G$, with 
$\mathcal S_G^0=\emptyset$.  Then the $i$th item added to $S_G$ is given by
\begin{align}
	f_i & = \arg\max_{f\in [m]/S_G^{i-1}}U\paranth{S_G^{i-1}\bigcup f}-U\paranth{S_G^{i-1}},~~~1\leq i \leq k\label{eq:greedy}.
\end{align}
For submodular functions, we have the following well known result.
\begin{thm}[\cite{nemhauser1978analysis}]
Let $U(\mathcal S)$ be a  non-decreasing submodular function.  Let $\mathcal S_G$ be the greedy maximizer of $U(\mathcal S)$ and let $\mathcal S_W$ be the maximizer of $U(\mathcal S)$ .  Then
\begin{align}
	   \frac{U(\mathcal S_G)}{U(\mathcal S_W)}&\geq 1-e^{-1}.
\end{align}
\end{thm}
Therefore, greedy solutions have performance guarantees for submodular maximization problems.  In addition, many times they can be computed very easily.  For these reasons, we will use a greedy approach for our problem.
 \subsection{Independent Events}
 We first consider the simplest case where all $E_i$ are independent.  Let $p_i=\mathbf P\paranth{E_i}$ and let $E_i^c$ denote the complement of $E_i$.  Because of the independence of the events, we can rewrite the objective function as 
\begin{align}
	U(\mathcal S) & = 1-\mathbf P\paranth{\bigcap_{i\in \mathcal S}E_i^c}\nonumber\\
	              & = 1- \prod_{i\in \mathcal S}{(1-p_i)}\label{eq:Uind}.
\end{align}

The optimization problem can then be written as

 \begin{align}\label{eq:opt_prob_min}
 	\max_{\mathcal S\subseteq [m], |\mathcal S|=k}~U(\mathcal S)&=\min_{\mathcal S\subseteq \mathcal [m], |\mathcal S|=k}~\prod_{i\in S}(1-p_i).
 \end{align}
The form of the objective function for independent events leads to a simple solution, given by the following theorem.
\begin{thm}\label{thm:Uind}
For a set of independent events $\mathcal E = \curly{E_1,E_2,...,E_m}$ let $p_i = \mathbf P(E_i)$ and without loss of generality assume
that $p_1\geq p_2\geq ...\geq p_m$.  Let the $k$ element subset $\mathcal S_W\subseteq [m]$ maximize $U(\mathcal S)$.  Then $\mathcal S_W$ consists of the indices with the largest $p_i$, i.e. $\mathcal S_W =[k]$.
\end{thm}
The proof of this result is obvious from the form of $U(\mathcal S)$ in equation \eqref{eq:Uind}. From Theorem \ref{thm:Uind} we see that if the events are independent, then one simply chooses the most likely events in order to maximize the probability that at least one event occurs.  This simple solution relies crucially upon the independence of the events.  Another useful property when the events are independent concerns the greedy solution.
\begin{lem}\label{thm:Uind_greedy}
 For a set of independent events $\mathcal E = \curly{E_1,E_2,...,E_m}$, let the $k$ element greedy maximizer of $U(\mathcal S)$ be $\mathcal S_G$.  Then $\mathcal S_W=\mathcal S_G$.
\end{lem}
In this case we have that the greedy solution is also an optimal solution.  Indeed, the independent events case is  almost trivial  to solve.  This also suggests when the events have low dependence, greedy solutions should perform well.

\subsection{Picking Winners and Log-Optimal Portfolios}
We now study the relationship between the picking winners portfolio and another well known portfolio known as a \emph{log-optimal portfolio}.
To do this, we consider a model where the items have a top-heavy payoff structure, similar to the picking winners problem.
We will let the random variable $X_i$ denote the return of the item associated with index $i$, and we assume
there are $m$ possible items. In particular, we will consider the case where there are two possible values for the returns $a$ and $b$ such that $0 < a < b$, and we will let $X_i(\omega) = b$ for $\omega \in E_i$, and $X_i(\omega) = a$ otherwise.  This can approximately model the returns
of a startup company which are huge if it exits, otherwise are negligible.  Because it can be hard to predict
the exact returns when a company exits, one can assume that they all have the same average value $b$ given no other
information.  In this way, the only factor differentiating companies is their probability of exiting.

  There are many different ways to construct a portfolio. The traditional Markowitz approach maximizes the mean portfolio return subject to an upper bound on its variance \citep{markowitz1952portfolio}. These portfolios are parametric because one must select the upper bound on the variance. Another approach is the log-optimal portfolio, where one constructs a portfolio which maximizes the expected natural logarithm of the portfolio return \citep{cover1991universal}.  In our analysis, we assume that investment decisions are binary for each item, i.e. an item is either in the portfolio or it is not.  We do not allow for continuous investment amounts for the items.
	Under this setting, a portfolio is a subset of $[m]$, as in the picking winners problem.
	We define the return of a portfolio $\mathcal S$ as $\sum\limits_{i \in \mathcal{S}} X_i / |\mathcal{S}|$.  The corresponding expected log return is
	\begin{align}
		V(\mathcal S)= \mathbf{E} \left[\ln \left( \sum\limits_{i \in \mathcal{S}} X_i / |\mathcal{S}|  \right) \right]. 
	\end{align}
	We constrain the portfolio to contain $k$ elements, as in the picking winners problem.
 Then the log-optimal portfolio is defined by the following optimization problem:
\begin{align}
	\max_{\mathcal S\subseteq [m], |\mathcal S| = k  } V(\mathcal S). \label{eq:log_optimal_problem}
\end{align}
There are no arbitrary parameters in the log-optimal portfolio formulation, unlike the Markowitz portfolio.  This allows us to directly compare the log-optimal  and picking winners portfolios.  Once again, we begin by considering the simple case where the events $\left\{ E_i \right\}_{i=1}^m$ are independent.   We have the following result:
\begin{thm}\label{thm:log_optimal_independent}
For a set of independent events $\mathcal E = \curly{E_1,E_2,...,E_m}$ let $p_i = \mathbf P(E_i)$ and without loss of generality assume
that $p_1\geq p_2\geq ...\geq p_m$.  Let the $k$ element subset $\mathcal S_L\subset [m]$ maximize $V(\mathcal S)$.  Then $\mathcal S_L$ consists of the indices with the largest $p_i$, i.e. $\mathcal S_L =[k]$.
\end{thm}
Comparing the above result and Theorem \ref{thm:Uind}, we see that when the events are independent the problem of determining a log-optimal portfolio reduces to the problem of finding a greedy solution to the picking winners problem.  The equivalence only relies on the assumptions that $0 < a < b$ and independence. In particular, we make no assumptions about the value of the marginal probabilities $\mathbf P (E_i)$.

While the independence assumption results in a clean equivalence between the log-optimal and picking winners
portfolios, in reality this assumption will be violated. However,  we are interested in examples where the probability of a particular item  winning is small. In this setting, we can quantify how much the two portfolio
types deviate from each other.
\begin{thm}\label{thm:log_optimal_small_dependence}
Let $\mathcal S_L$ denote a log-optimal portfolio, i.e. it is optimal for \eqref{eq:log_optimal_problem}, and let $\mathcal S_W$ denote a portfolio that is optimal for the picking winners problem. Let $\mathcal G_l$ denote the collection of all subsets of $[m]$ with cardinality $l$. Suppose there exists $\lambda \in [0, 1]$ and $p \in \left(0, \frac{1}{k}\right]$ such that for all $l \in [k]$, and for all $T\in \mathcal G_l$ the following holds 
\begin{align}
(1-\lambda) p^l (1-p)^{k-l} \leq \mathbf P \left(  \left(\bigcap\limits_{i \in T} E_i \right) \bigcap \left(  \bigcap\limits_{j\in\mathcal S \setminus T} F_j^c \right) \right) \leq (1+\lambda) p^l (1-p)^{k-l} . \label{eq:assumption}
\end{align}  
Then it follows that 
\begin{align*}
\frac{U(S_W)-U(S_L)}{U(S_W)}  \leq \frac{2\zeta(3)\lambda kp(1-p)}
{\ln\left(1 + \frac{b-a}{ka}\right)(1-\lambda)(1-(1-p)^k)}.
\end{align*}
where $\zeta(s)$ is the Riemann zeta function evaluated at $s$, i.e. $\zeta(s) = \sum\limits_{n=1}^s \frac{1}{n^s}$.
\end{thm}

We will begin by interpreting the assumptions for Theorem \ref{thm:log_optimal_small_dependence}. We begin by considering \eqref{eq:assumption}. This assumption effectively places restrictions on the dependency between events. To gain some intution, consider the case where all of the events have the same probability, and let $p_i = p$ for all $i \in [m]$. Then it is clear that $\lambda$ is a measure of the dependency between events. In particular, when the events are independent we can choose $\lambda =0$, and as the events become more correlated we expect that we will have to choose a larger $\lambda$ for equation \eqref{eq:assumption} to hold. Additionally, Theorem \ref{thm:log_optimal_small_dependence} does not require the probability of all the events to be the same, it just requires there to be some $p \in \left(0, \frac{1}{k} \right]$ and some $\lambda \in [0, 1]$ where \eqref{eq:assumption} holds. If the items have very different values for $p_i$, then Theorem \ref{thm:log_optimal_small_dependence} might produce a very weak bound. If all the events of interest have an equal probability of winning, and once again we we let let $p = p_i$ for all $i \in [m]$, then the assumption $p \in \left(0, \frac{1}{k} \right]$ implies that the expected number of winning items for any portfolio of size $k$ is one or less. For our applications of interest, the winning probabilities are somewhat small and thus this seems like a reasonable assumption. 

The main insight of Theorem \ref{thm:log_optimal_small_dependence} is that for our particular applications of interest, a log-optimal portfolio must also be a good solution to the picking winners problem. Throughout this discussion, suppose that the conditions in the theorem hold. As an example,  assume that $\lambda = 0.5$, $k= 10$, and $p=0.01$.  This assumes that the  probability of an individual item winning is low and  allows from some correlation.  For the possible values of the returns, let us assume $b = \$10^9$ and $a=\$1$ (these are the approximate values for a very successful exit, such as for Facebook).  
Then applying Theorem \ref{thm:log_optimal_small_dependence} we have that the percent difference in the objective function (probability of at least one winner)
for a winning and log-optimal portfolio is less than $27\%$.  This example illustrates that under these conditions, which are relevant for investing in startup companies, the optimal portfolio for the picking winners problem will be similar in performance to a log-optimal portfolio.  



\subsection{Portfolio Construction for Brownian Motion Model}\label{sec:exit_prob}
We now discuss portfolio construction for our Brownian motion model.  Define $E_i$ as the event that company $i$ reaches an exit sometime in the future.  We wish to build a portfolio of $k$ companies to maximize the probability that at least one company exits. To do this, we must be able to evaluate the exit probability. In our models this can easily be done. Recall that for a set of companies $\mathcal S$, we define $U(\mathcal S) = \mathbf P\paranth{\bigcup_{i\in\mathcal S} E_i}$. We now show how to calculate the exit probabilities.

If we assume that the performance of the companies is independent, then all we need to calculate is $p_i = \mathbf P(E_i)$ for each company $i\in\mathcal S$ to construct the portfolio.  Recall that this is the probability
that the company's Brownian motion hits level $7\Delta$ in finite time.
We assume we are given $\mu_{i}(t)$ and $\sigma_i(t)$ for company $i$.
Then, from  \eqref{eq:cdf} the exit probability is 
\begin{align}
p_i & = \lim_{t\to \infty} F_{0}\paranth{t; 0, \mu_i(t),\sigma_i(t),7\Delta}  \label{eq:pi}.
\end{align}
Additionally, recall that we are uncertain in the future values of $\beta_y$.  This causes uncertainty in the drift which results in the companies no longer being independent.  To see  why this is the case, assume we are given the parameter vector $\beta_{y-1}$.  This will be the situation when we train on past data up to year $y-1$ and try to build a portfolio for companies founded in year $y$.  Therefore, we need to know $\beta_y =[\beta_{y1},\beta_{y2},...\beta_{yM}]$.  Recall that $\beta_{yi}$ is normally distributed with mean 
$\beta_{y-1,i}$ and standard deviation $\delta_i$.  This results in random drift coefficients for all companies founded in year $y$.  We do not know the exact value of the company drifts, but we do know that they will all be affected the same way by the actual realization of $\beta_y$.  This is how the uncertainty creates a positive correlation between the companies.  To calculate $U(\mathcal S)$, we simply average over the uncertainty in $\beta_y$.  Using equations \eqref{eq:Uind} and \eqref{eq:pi} we have 
\begin{align}
U(\mathcal S) & = \mathbf E_{\beta_y}\bracket{\mathbf P\paranth{\bigcup_{i\in\mathcal S} E_i\Bigg |\beta_y}}\nonumber\\
& = 1-\mathbf E_{\beta_y}\bracket{\prod_{i\in \mathcal S}(1-p_i)}\nonumber\\
&=1-\mathbf E_{\beta_y}\bracket{\prod_{i\in \mathcal S} (1- \lim_{t\to \infty} F_{0}\paranth{t;t_{i_0}, \mu_i(t),\sigma_i(t),7\Delta})}.\label{eq:pi_correlated}
\end{align}
Because $\beta_y$ is jointly normal and it is simple to evaluate $F_{0}(\cdot)$, this expectation can be easily done using Monte Carlo integration.

For our model, because the number of startup companies that we are choosing from is quite large, we create our portfolio using the greedy approach.  This is computationally feasible because in practice the portfolio will not be very large (typical venture capital firms manage no more than a few hundred companies) and it is easy to evaluate $U(\mathcal S)$ for a set $\mathcal S$. We consider the performance of portfolios built using our model in Section \ref{sec:portfoliosection}.

\section{Model Validation with Picking Winners Portfolios}
\label{sec:portfoliosection}

While we tested our models for many different years, we report the results for test years 2011 and 2012.  We use these years because it gives a sufficient amount of time for the companies to achieve an exit (we collected the data to conduct this analysis in late 2016). For each test year, the test set is equal to the companies with baseline year (earliest funding year) equal to that year, and all other companies used in model estimation form the training set.  The features used to calculate the exit probabilities of the test companies are built using only data that would have been available at the time of the companies' earliest funding round. 

We use the greedy portfolio construction method from equation \eqref{eq:greedy}. We build portfolios under two assumptions, one that companies are dependent and the other that they are independent. The difference in the two assumptions is how we average over uncertainty in the drift parameters. For the dependent assumption, we do the averaging for all companies jointly, as shown in equation \eqref{eq:pi_correlated}. For the independent assumption, we average equation \eqref{eq:pi} over the uncertainty for each company individually, and then use the resulting exit probabilities to construct the portfolio. The averaging is done using Monte Carlo integration. Specifically, for every $\beta$ sample, we generate a random draw from a normal distribution for each element of the $\beta$ vector for the testing year, with mean given by $\beta_{2010}$ or $\beta_{2011}$ (depending on the testing year) and standard deviation given by the corresponding element in $\delta$. We then average the relevant function over these random variates.

\begin{figure}
	\includegraphics[scale=0.43]{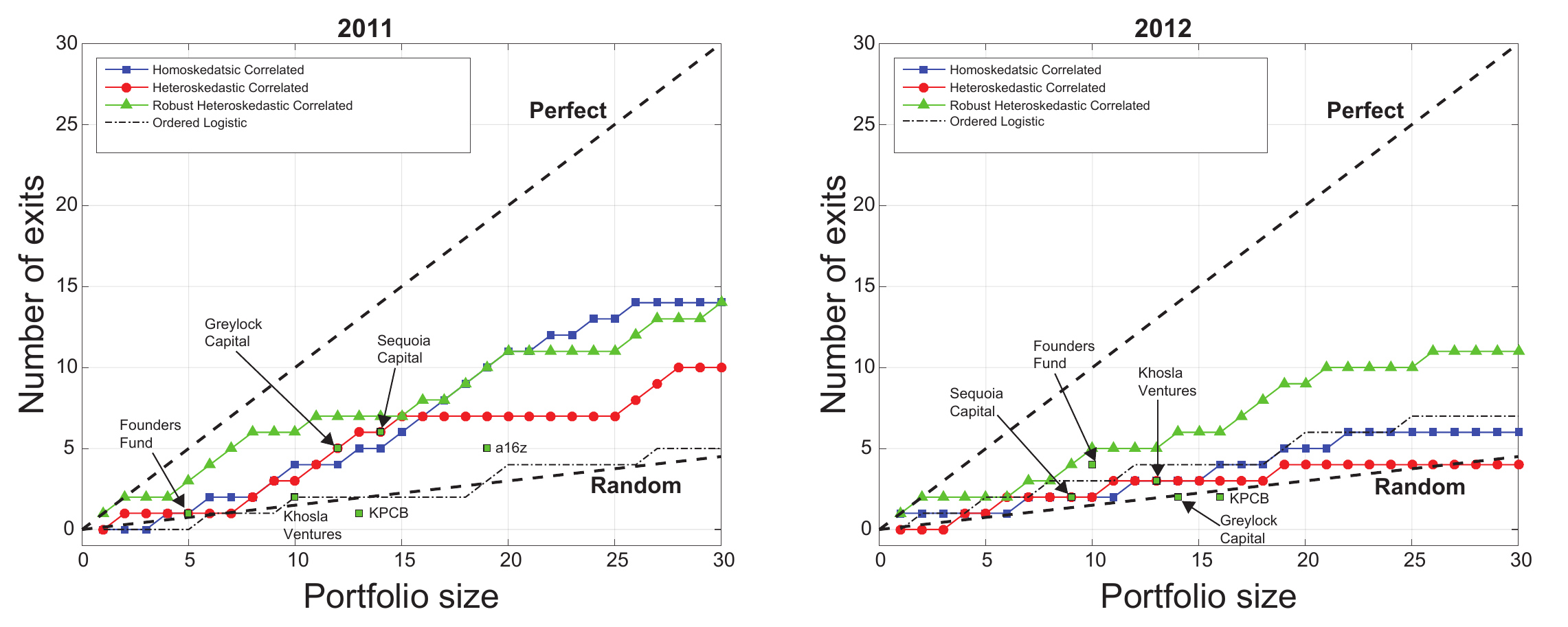}
	\caption{Performance curves for different model portfolios constructed using companies with seed or series A funding in 2011 (left) and 2012 (right). Also shown are the performances of venture capital firms Greylock Partners (Greylock), Founders Fund (Founders), Kleiner Perkins Caufield \& Byers (KPCB), Sequoia Capital (Sequoia), and Khosla Ventures (Khosla).  For reference the performance curves for a random portfolio and a perfect portfolio are also shown.}
	\label{fig:performancegraph}
\end{figure}

We construct portfolios for the models detailed in Section \ref{sec:bayesmodelsection} with a maximum of 30 companies, which is a reasonable size for a typical venture capital firm.  We also benchmark our model against an ordered logistic regression model using our features and labels corresponding to the largest round that a company achieved before the test year.  To visualize the performance of our portfolios, we plot the portfolio size versus the number of companies  in the portfolio that exit for select models  in Figure \ref{fig:performancegraph}.  We refer to this as a portfolio performance curve. It is similar to a receiver operating characteristic (ROC) curve used in machine learning to evaluate the performance of binary classifiers. 

We consider different types of portfolios. Ordered logistic regression is our benchmark model as discussed earlier.  Correlated correspond to our Brownian motion model with the dependence assumption. Perfect corresponds to the portfolio where every company exits. Random corresponds to portfolios where the number of exits is the portfolio size multiplied by the fraction of exits in the test set. This corresponds to picking a portfolio randomly.  We also show the performance points for top venture capital firms in these years. The coordinates of these points are the number of test companies that the venture capital firm invests in and the number of these companies that exit before 2017.  


We find that the  the robust heteroskedastic model has the best performance over both years.  It obtains six exits with a ten company portfolio in 2011, giving an exit rate of 60\%.  This is higher than the top venture capital firms shown in Figure \ref{fig:performancegraph}.  In contrast, the homoskedastic and heteroskedastic models do not consistently perform well.  Also, the ordered logistic model has terrible performance in 2011, but is as good as the homoskedastic model in 2012.  

We list the companies for robust heteroskedastic model portfolios in Tables \ref{table:m3porfolios2011} and \ref{table:m3porfolios2012}.  As can be seen, all exits found are acquisitions, which are more common than IPOs.  Some of the acquisitions were significant.  For instance, the mobile gaming company Funzio in the 2011 portfolio was acquired for \$210 million by GREE in 2012 \citep{funzio}.     We show the portfolios  for all  models and years in  Appendix \ref{app:portfoliotablesall}.

\begin{table}
	\centering
	\caption{The top companies in our 2011 portfolio constructed with greedy optimization and the \textbf{robust heteroskedastic correlated} model.
	Shown are the company name, the highest funding round achieved, the predicted exit probability, and the greedy objective value (probability of at least one exit) when the company is added rounded to the nearest hundredth.
Companies which exited are highlighted in bold.}
	\label{table:m3porfolios2011}
	\begin{tabular}{|c|c|c|c|}
		\hline
		\textbf{2011 Company} & \textbf{Highest funding round} & \textbf{Exit probability}&\textbf{ Objective value}\\	\hline
                        \textbf{Funzio} & Acquisition & 0.54 & 0.54 \\   \hline
                        \textbf{Longboard Media} & Acquisition & 0.43 & 0.74 \\   \hline
                        AirTouch Communications & A & 0.49 & 0.86 \\   \hline
                        Sequent & B & 0.47 & 0.92 \\    \hline
                        \textbf{Bitzer Mobile} & Acquisition & 0.47 & 0.95 \\   \hline
                        \textbf{SHIFT} & Acquisition & 0.44 & 0.96  \\   \hline
                        \textbf{Livestar} & Acquisition & 0.38 & 0.97 \\   \hline
                        \textbf{Jybe} & Acquisition & 0.4 & 0.98 \\   \hline
                        Whitetruffle & Seed & 0.4 & 0.99  \\   \hline
                        Adteractive & A & 0.27 & 0.99 \\   \hline
	\end{tabular}
\end{table}

\begin{table}
	\centering
	\caption{The top companies in our 2012 portfolio constructed with greedy optimization and the \textbf{robust heteroskedastic correlated} model.
		Shown are the company name, the highest funding round achieved, the predicted exit probability, and the greedy objective value (probability of at least one exit) when the company is added rounded to the nearest hundredth. Companies which exited are highlighted in bold.}
	\label{table:m3porfolios2012}
	\begin{tabular}{|c|c|c|c|}
		\hline
		\textbf{2012 Company} & \textbf{Highest funding round} & \textbf{Exit probability}&\textbf{ Objective value}\\	\hline
                        \textbf{Struq} & Acquisition & 0.76 & 0.76 \\   \hline
                        \textbf{Metaresolver} & Acquisition & 0.73 & 0.93 \\   \hline
                        Trv & C & 0.67 & 0.97 \\   \hline
                        Glossi  Inc & Seed & 0.75 & 0.99 \\   \hline
                        Boomerang & Seed & 0.71 & 0.99 \\   \hline
                        Alicanto & A & 0.3 & 1.0 \\   \hline
                        \textbf{SnappyTV} & Acquisition & 0.25 & 1.0 \\   \hline
                        AVOS Systems & A & 0.17 & 1.0 \\   \hline
                        \textbf{Adept Cloud} & Acquisition & 0.17 & 1.0 \\   \hline
                        \textbf{Runa} & Acquisition & 0.21 & 1.0 \\   \hline
	\end{tabular}
\end{table}

\section{Conclusion}\label{sec:conclusion}
We have presented a Bayesian modeling framework for evaluating the quality of startup companies
using online data. We first analyzed features constructed from our dataset to gain an empirical understanding 
of which factors influence company success. Then, we introduced a Brownian motion model for the evolution of a
startup company's funding rounds to calculate
the relevant probabilities of exiting. 
From there, we developed four variants
of the Brownian motion model using either different likelihood or prior distribution specifications. We estimated the four models
on data for thousands of companies. Because we focused on a Bayesian modeling framework, we 
were able to use our parameter estimations to determine which features are associated with successful companies
with statistical significance. Furthermore, our use of hierarchical Bayesian modeling to estimate a sector-specific model
allowed us to gain insight into how the factors of success differ for companies in different industries.

In addition to our framework for evaluating the quality of startups, we have presented the picking winners framework for constructing
portfolios of items which either have huge returns 
or very low returns.  We formulated the problem
as one of maximizing the probability of at least one item achieving a large return, or winning. The submodularity of this
probability allowed us to efficiently construct
portfolios using a greedy approach. We applied our portfolio construction
method to startup companies, which have this top-heavy return structure. To calculate
the relevant probabilities of winning, we used our estimated Brownian motion model. We constructed portfolios
with our approach which outperform top venture capital firms.
The results show that our modeling and portfolio
construction method are effective and 
provide a quantitative methodology for evaluating the quality of startup companies.

There are several future steps for this work.  One
concerns the manner in which company exits are treated.
Acquisitions and IPOs are equivalent in our model, but
in reality IPOs are much more rare and can result
in a much higher payoff.  Our model could be modified
to distinguish between IPOs and acquisitions, or it could
incorporate the monetary value of the IPO or acquisition.
For IPOs this valuation data is public, but acquisition
data is generally private and may be difficult to obtain. 

Another direction concerns the features used in our model.
While we included many relevant features, there could be other useful
features which have predictive power.  For instance, the value of
different macro-economic indicators in the founding year of a company
may have impact on its exit probability.  In addition to the the types
of features used, one could also explore the way in which they are incorporated
into the model.  We used a simple linear mapping between the features
and the drift and diffusion.  However, more complex 
models which use non-linear mappings may improve performance.

\begin{APPENDICES}
	\section{Proofs}
	\subsection{Proof of Theorem \ref{thm:nphard}}
	We show that maximizing $U(\mathcal S)$ is NP-hard by reducing it to the maximum coverage problem.
	In this problem one is given a set $\mathcal U$ of $n$ elements and a collection $\mathcal E =\curly{E_i}_{i=1}^N$ of
	$N$ subsets of $\mathcal U$ such that $\bigcup_{E\in\mathcal E}E=\mathcal U$.   The goal is to select $M$ sets from $\mathcal E$ such that their union has maximum cardinality.  This is known to
	be an NP-hard optimization problem.  To show that this is an instance of maximizing $U(\mathcal S)$ we assume
	that the sample space $\Omega$ is countable and finite with $R$ elements.  We also assume that each element $\omega\in\Omega$
	has equal probability, i.e. $\mathbf P(\omega)=R^{-1}$. Let $\mathcal F$ be the $\sigma$-algebra of $\Omega$.   For any set $\mathcal S\in\mathcal F$, we can write $U(\mathcal S) = R^{-1}|\bigcup_{\omega\in\mathcal S}\omega|$.
	Then we have
	\begin{align*}
	\max_{\mathcal S\subseteq \mathcal E, |\mathcal S|=M} U(\mathcal S) & = \max_{\mathcal S\subseteq \mathcal E, |\mathcal S|=M}R^{-1}\left|\bigcup_{\omega\in\mathcal S}\omega\right|\\
	&=\max_{\mathcal S\subseteq \mathcal E, |\mathcal S|=M}\left|\bigcup_{\omega\in\mathcal S}\omega\right|.
	\end{align*}
	Therefore, maximizing $U(\mathcal S)$ is equivalent to the maximum coverage problem.
	\subsection{Proof of Lemma \ref{thm:submodular}}
	The function $U(\mathcal S)$ is non-negative and non-decreasing because it is the probability of a set of events.  
	We must show that it is also submodular.  A submodular function $f$ satisfies 
	\begin{align}
	f\paranth{\mathcal S \bigcup v} - f\paranth{\mathcal S } &\geq f\paranth{\mathcal T \bigcup v}  - f\paranth{\mathcal T} 
	\end{align}
	for all elements $v$ and pairs of sets $\mathcal S$ and $\mathcal T$ such that
	$\mathcal S\subseteq \mathcal T$.
	We show that the function $U(\mathcal S)$ is submodular as follows.  We let the $\sigma$-algebra of the probability space be $\mathcal F$.   Consider sets $\mathcal S,\mathcal T, v\in\mathcal F$ such that $\mathcal S\subseteq \mathcal T$.
	We can write $v=v_\mathcal S \bigcup v_\mathcal T\bigcup v_0$ where we define $v_\mathcal S = v\bigcap \mathcal S$, $v_\mathcal T = v\bigcap \mathcal T\bigcap \mathcal S^c$, and $v_0 = v\bigcap \mathcal T^c$.  Then we have
	\begin{align}
	U\paranth{\mathcal T\bigcup v}-U\paranth{\mathcal T}&= \mathbf P\paranth{v_0}\nonumber
	\end{align}
	and
	\begin{align}
	U\paranth{\mathcal S\bigcup v}-U\paranth{\mathcal S}&= \mathbf P\paranth{v_\mathcal T \bigcup v_0}\nonumber\\
	&\geq \mathbf P\paranth{v_0}\nonumber\\
	&\geq 	U\paranth{\mathcal T\bigcup v}-U\paranth{\mathcal T},\nonumber
	\end{align}
	thus showing that $U(\mathcal S)$ satisfies the submodularity condition.

	\subsection{Proof of Theorem \ref{thm:log_optimal_independent}}\label{sec:log_optim_independent}
	We will begin by defining some notation. For $0 \leq l \leq k$, let $\mathcal G_l(\mathcal S)$ denote the collection of all subsets of $\mathcal S$ with cardinality $l$. Additionally, for $0 \leq q \leq k$, we also define the events $W_q(\mathcal S)$ and $Y_q(\mathcal S)$ as follows 
	\begin{align*}
	W_q(\mathcal S) &= \bigcup_{T \in \mathcal G_q(\mathcal S)} \left( \left( \bigcap\limits_{i \in T} E_i \right) \bigcap \left( \bigcap_{j \in \mathcal{S} \setminus T}E_j^c  \right)    \right) \\ \\
	Y_q(\mathcal S) &= \bigcup\limits_{t=q}^k W_t(\mathcal S). \quad 
	\end{align*}
	In particular, $W_q(\mathcal S)$ is the event where exactly $q$ of the event $E_i$ for $i \in \mathcal S$ occur. Naturally, $Y_q(\mathcal S)$ is the event where $q$ or more of the events $E_i$ for $i \in \mathcal S$ occur. Using this notation, when $\mathcal S$ is of cardinality $k$ we get the following
	\begin{align}
	V(\mathcal S) &= \sum\limits_{q=0}^k \mathbf P(W_q(\mathcal S)) \ln\left( \frac{qb+(k-q)a}{k} \right) \nonumber\\\nonumber \\
	&= \sum\limits_{q=0}^{k-1} \left( \mathbf P(Y_q(\mathcal S)) - \mathbf P(Y_{q+1}(\mathcal S))\right) \ln\left( \frac{qb+(k-q)a}{k} \right) + \mathbf P(Y_k) \ln \left(\frac{kb}{k} \right) \nonumber\\\nonumber \\
	&= \ln(a) + \sum\limits_{q=1}^k \mathbf P \left(Y_q(\mathcal S) \right) \ln\left( 1+ \frac{b-a}{(q-1)b + (k-q+1)a} \right) .\label{eq:third} 
	\end{align}
	The second line above follows from the definition of $Y_q(\mathcal S) $ and $W_q(\mathcal S) $, and the third line follows from the basic properites of a logarithm. We now will conclude by showing that $\mathcal S = [k]$ maximizes every term in the sum given in \eqref{eq:third}, and thus it is maximizes $V(\mathcal S)$. Now, because $b >a$ by assumption, we have that 
	\begin{align*}
	\ln\left( 1+ \frac{b-a}{(q-1)b + (k-q+1)a} \right) > 0.\quad 
	\end{align*}
	for all $q \in [k]$. Due to the positivity of this value, maximizing every term in the sum given in \eqref{eq:third} reduces to the problem of maximizing $\mathbf P \left( Y_q(\mathcal S ) \right)$. Finally, we show that that $\mathcal S = [k]$ maximizes $\mathbf P (Y_q (\mathcal S ))$ for all $q \in [k]$. If $[k]$ is a maximizer for $\mathbf P (Y_q(\mathcal S))$, then we are done.  We define $S_L = [k]$ and for purpose of contradiction, we suppose that there exists an $\mathcal S^\dagger$ of cardinality $k$ such that $\mathcal S^\dagger \neq S_L$, and $\mathbf P (Y_q (\mathcal S^\dagger ) )> \mathbf P (Y_q (S_L ))$. Therefore, we know
	there exists a $r \in \mathcal S^\dagger$ and a $t \in \mathcal S_L$ such that $r\notin \mathcal S_L$, 
	$t\notin \mathcal S^\dagger$, and	$p_t \geq p_r$, otherwise we would have $\mathcal S^\dagger = S_L$. Then from the definition of $Y_q(\mathcal{S})$, we have that
	\begin{align*}
	\mathbf P (Y_q (\mathcal S^\dagger) ) &= p_r \mathbf P \left( Y_q(\mathcal S^\dagger) \; | \; E_r  \right) + (1-p_r) \mathbf P \left( Y_q(\mathcal S^\dagger) \; | \; E_r^c  \right) \\
	&= \mathbf P \left( Y_q(\mathcal S^\dagger) \; | \; E_r^c  \right) +p_r\left(  \mathbf P \left( Y_q(\mathcal S^\dagger) \; | \; E_r  \right)-\mathbf P \left( Y_q(\mathcal S^\dagger) \; | \; E_r^c  \right)  \right ) \\
	&= \mathbf{P}( Y_q(\mathcal S^\dagger \setminus \{r\})) + p_r \left( \mathbf P \left( Y_{q-1}(\mathcal S^\dagger \setminus \{r\}) \right) - \mathbf P \left( Y_q(\mathcal S^\dagger \setminus \{r\}) \right) \right). \quad 
	\end{align*} 
	By a similar logic, we have that
	\begin{align*}
	\mathbf P (Y_q(\mathcal S^\dagger \cup \{ t\} \setminus \{r\}) ) &= \mathbf{P}( Y_q(\mathcal S^\dagger \setminus \{r\})) + p_t \left( \mathbf P \left( Y_{q-1}(\mathcal S^\dagger \setminus \{r\}) \right) - \mathbf P \left( Y_q(\mathcal S^\dagger \setminus \{r\}) \right) \right). \quad
	\end{align*}
	Finally, $Y_{q-1}(\mathcal S^\dagger \setminus \{r\}) \supseteq  Y_q(\mathcal S^\dagger \setminus \{r\}) $ and thus it follows that $ \mathbf P (Y_q (\mathcal S^\dagger) ) \leq \mathbf P (Y_q(\mathcal S^\dagger \cup \{ t\} \setminus \{r\}) )$. By repeating this argument for all elements of $\mathcal S^\dagger$ that are not in $S_L$, we can arrive at the conclusion $\mathbf P (Y_q (\mathcal S^\dagger ) ) \leq \mathbf P (Y_q (S_L ))$. However, this is a contradiction, and thus $S_L$ is a maximizer for $\mathbf P (Y_q (\mathcal S ))$.

	\subsection{Proof of Theorem \ref{thm:log_optimal_small_dependence}}
	We will begin in a way that is very similar to the proof in section \ref{sec:log_optim_independent}. In particular, we begin by defining $W_q(\mathcal S)$ and $Y_q(\mathcal S)$ for $l \in [k]$ as follows 
	\begin{align*}
	W_q(\mathcal S) &= \bigcup_{T \in \mathcal G_q(\mathcal S)} \left( \left( \bigcap\limits_{i \in T} E_i \right) \bigcap \left( \bigcap_{j \in \mathcal{S} \setminus T}E_j^c  \right)    \right) \\ \\
	Y_q(\mathcal S) &= \bigcup\limits_{t=q}^k W_t(\mathcal S). \quad 
	\end{align*}
	Now, let $S_k$ denote a binomial random variable with parameters $k$ and $p$. Then, from the assumptions given in the statement of the theorem, we have that for the log-optimal portfolio $S_L$,
	\begin{align*}
	\mathbf{P} \left(Y_j \left(\mathcal S_L \right) \right) &= \sum\limits_{i=j}^k \sum\limits_{T \in \mathcal{G}_i(\mathcal S_L)} \mathbf{P} \left( \left(\bigcap\limits_{E \in T} E \right) \bigcap \left(  \bigcap\limits_{F\in\mathcal S_L \setminus T} F^c \right)  \right) \\
	&\leq \sum\limits_{i=j}^k \sum\limits_{T \in \mathcal{G}_i(\mathcal S_L)} \left(1 + \lambda \right) p^i (1-p)^{k-i} \\
	&= (1+ \lambda) \sum\limits_{i=j}^k {{k}\choose{i}} p^i (1-p)^{k-i} \\
	&= (1+\lambda) \mathbf{P}\left( S_k \geq j  \right). \quad 
	\end{align*}
	for all $j \in [k]$. By the same logic, we also get that for the picking winners portfolio $S_W$,
	\begin{align*}
	\mathbf{P} \left(Y_j \left(\mathcal S_W \right) \right) \geq (1-\lambda) \mathbf{P}\left( S_k \geq j  \right) \quad . 
	\end{align*}
	Then, using equation \eqref{eq:third}, the inequalities given above, and the fact that $\mathbf P (Y_1(\mathcal S)) = U(\mathcal S)$, we obtain
	\begin{align*}
	V(\mathcal S_L) - V(\mathcal S_W) \leq \left(U(S_L)-U(S_W)\right) \ln\left(1 + \frac{b-a}{ka}\right) + 2\lambda \sum\limits_{j=2}^k \mathbf P (S_k \geq j) \ln \left(  1 + \frac{b-a}{ka + (j-1)(b-a)} \right). \quad 
	\end{align*}
	Now, using Chebyshev's inequality, the variance of a binomial random variable, and the above inequality we have that
	\begin{align*}
	V(\mathcal S_L) - V(\mathcal  S_W) &\leq \left(U(S_L)-U(S_W)\right) \ln\left(1 + \frac{b-a}{ka}\right) + 2\lambda \sum\limits_{j=2}^k\mathbf P \left( |S_n-kp| \geq j-kp  \right) \ln \left(  1 + \frac{b-a}{ka + (j-1)(b-a)} \right) \\ \\
	&\leq \left(U(S_L)-U(S_W)\right) \ln\left(1 + \frac{b-a}{ka}\right) + 2\lambda \sum\limits_{j=2}^k \frac{kp(1-p)}{(j-kp)^2} \ln \left(  1 + \frac{b-a}{ka + (j-1)(b-a)} \right). \quad
	\end{align*}
	Now, recall that $\ln(1+x) \leq x$ for $x > -1$. Therefore
	\begin{align*}
	V(\mathcal S_L) - V(\mathcal  S_W) &\leq \left(U(S_L)-U(S_W)\right) \ln\left(1 + \frac{b-a}{ka}\right) + 2\lambda \sum\limits_{j=2}^k \frac{kp(1-p)}{(j-kp)^2} \frac{1}{j-1} \\ \\
	&\leq  \left(U(S_L)-U(S_W)\right) \ln\left(1 + \frac{b-a}{ka}\right) + 2\lambda kp(1-p)\sum\limits_{j=2}^k \frac{1}{(j-1)^3} 
	\end{align*}
	where for the second inequality we used the assumption $p\in \left(0, \frac{1}{k} \right]$. Some
	further analysis gives 
	\begin{align}
	V(\mathcal S_L) - V(\mathcal S_W) &\leq  c \ln\left(1 + \frac{b-a}{ka}\right) + 2\lambda kp(1-p)\sum\limits_{j=2}^k \frac{1}{(j-1)^3}\nonumber \\ \nonumber\\
	&\leq \left(U(S_L)-U(S_W)\right) \ln\left(1 + \frac{b-a}{ka}\right) + 2\lambda kp(1-p)\sum\limits_{j=1}^\infty \frac{1}{j^3}\nonumber \\ \nonumber\\
	&\leq \left(U(S_L)-U(S_W)\right) \ln\left(1 + \frac{b-a}{ka}\right) + 2\lambda kp(1-p)\zeta (3).\label{eq:bound}
	\end{align}
	Now, because $S_L$ is log-optimal, we have that $ V(\mathcal S_L) - V(\mathcal  S_W)\geq 0$.
	Using this relationship we can rearrange equation \eqref{eq:bound} to obtain
		\begin{align}
	 \left(U(S_W)-U(S_L)\right) \ln\left(1 + \frac{b-a}{ka}\right)& \leq 2\lambda kp(1-p)\zeta(3)\nonumber \\
	     &\nonumber \\ 
	  	  U(S_W)-U(S_L)  &\leq \frac{2\lambda kp(1-p)\zeta(3)}{\ln\left(1 + \frac{b-a}{ka}\right)}.\label{eq:bound2}
	\end{align}
	We next lower bound $U(S_W)$ using the assumptions of Theormem \ref{thm:log_optimal_small_dependence} and basic propertiesof the binomial distribution to obtain
    \begin{align}
			U(S_W) &= 	\mathbf{P} \left(Y_j \left(\mathcal S_W \right) \right)\nonumber\nonumber \\
			& = (1-\lambda) \mathbf{P}\left( S_k \geq 1  \right)\nonumber\nonumber \\
			&\geq  (1-\lambda)(1-(1-p)^k).\label{eq:Ubound}
	\end{align}
	Combining this lower bound with equation \eqref{eq:bound2} we obtain our final result.
			\begin{align*}
	\frac{U(S_W)-U(S_L)}{U(S_W)}  \leq \frac{2\lambda kp(1-p)\zeta(3)}
	                                {\ln\left(1 + \frac{b-a}{ka}\right)(1-\lambda)(1-(1-p)^k)}.
	\end{align*}

	\section{Sector Names}\label{sec:sector_names}
	The sector names that we used from the Crunchbase database are:
\emph{3d printing,	advertising,
analytics,
animation,
apps,
artificial intelligence,
automotive,
autonomous vehicles,
big data,
bioinformatics,
biotechnology,
bitcoin,
business intelligence,
cloud computing,
computer,
computer vision,
dating,
developer apis,
e-commerce,
e-learning,
edtech,
education,
Facebook,
fantasy sports,
fashion,
fintech,
finance,
financial services,
fitness,
gpu,
hardware,
health care,
health diagnostics,
hospital,
insurance,
internet,
internet of things,
iOS,
lifestyle,
logistics,
machine learning,
medical,
medical device,
messaging,
mobile,
nanotechnology,
network security,
open source,
personal health,
pet,
photo sharing,
renewable energy,
ride sharing,
robotics,
search engine,
social media,
social network,
software,
solar,
sports,
transportation,
video games,
virtual reality},
and \emph{virtualization}.

	
	\section{Top School Names}\label{app:schools}
	The schools used for the top school feature are:
\emph{Berkeley,
Brown,
California Institute of Technology,
Carnegie Mellon,
Columbia,
Cornell,
Dartmouth,
Duke,
Harvard,
Johns Hopkins,
Massachusetts Institute of Technology,
Northwestern,
Princeton,
Stanford,%
University of Chicago,
University of Pennsylvania,
Wharton,}
 and \emph{Yale.}

\section{Details of the MCMC Sampler}
\label{app:samplingdetails}

We use a blocked Metropolis-within-Gibbs scheme to sample from the posterior distribution of the model parameters. In Appendix \ref{sec:metropolisstep}, we describe the details of the Metropolis step used to sample the parameters. In the following sections of this appendix, we describe how we sample the parameters of each of the four models described in Section \ref{sec:bayesmodelsection}. Note that in the following sections we use $\Theta$ to describe the set of model parameters and the notation $\Theta_{-\alpha}$ to describe the set of model parameters excluding parameter $\alpha$.

\subsection{Details of the Metropolis Step}
\label{sec:metropolisstep}

Let $\alpha_i$ denote the $i$th sample of parameter block $\alpha$. In each step of the random walk Metropolis-Hastings, the proposal for sample $(i + 1)$ is drawn from a normal distribution with mean $\alpha_i$ and variance $\epsilon^2$. In the case where the parameter block $\alpha$ is a vector, we use a multivariate normal proposal distribution with mean $\alpha_i$ and diagonal covariance matrix equal to $\epsilon^2\mathbb{\textbf{I}}$. 

\cite{graves2011automatic} found that tuning the step size of MCMC algorithms helps with keeping a high acceptance probability during the sampling. Therefore, we adjust the step size $\epsilon$ based on the acceptance rate of previous samples according to the schedule proposed by the default implementation of the PyMC3 framework \citep{pymc3ref} as follows:
\begin{enumerate}
\item Initialize the step size to 1.
\item Every 100 samples, do the following:
\begin{itemize}
\item If the acceptance rate is less than .05, multiply $\epsilon$ by .5
\item If the acceptance rate is less than .2, multiply $\epsilon$ by .9
\item If the acceptance rate is greater than .5, multiply $\epsilon$ by 1.1
 \item If the acceptance rate is greater than .75, multiply $\epsilon$ by 2
 \item If the acceptance rate is greater than .95, multiply $\epsilon$ by 10
 \item Else do not change $\epsilon$
\end{itemize}
\end{enumerate}

\subsection{Homoskedastic Model Parameters}
\label{sec:homoskedasticsample}

The set of model parameters in the homoskedastic model described in Section \ref{sec:homoskedasticmodel} is $\Theta = $\{$\beta_y$ \text{for all $y$ in our data}, $\sigma_0^2$, $\delta$, $\nu$, $\tau$\}. For this model, we set the hyperparameters to be $\mu_\beta = 0$, $\sigma^2_\beta = 100$, $a_{\sigma_0^2} = 3$, $b_{\sigma_0^2} = 1$, $\lambda_0 = 1$, $a_\nu = 0$, $b_\nu = 50$, $a_\tau = -10$, and $b_\tau = 10$. In general, we choose our hyperpriors to be flat and uninformative. For the priors on $\nu$ and $\tau$, we incorporate information from our data analysis in Section \ref{sec:data}. Specifically, our data showed that no companies ever exit after 50 years, so we set the upper bound of $\nu$ to be 50. Also, from our data analysis, we can reasonably expect that company performance has a half life of less than 10 years, so we bound $\log(\tau)$ between -10 and 10.

We note that for this model, the posterior cannot be written in closed form for any of the model parameters due to lack of conjugacy between the data likelihood $\mathbb{P}(\textbf{T}, \textbf{X} | \Theta)$ and the parameter priors. Therefore, we sample each parameter block $\alpha \in \Theta$ using a random walk Metropolis step according to the rules in Appendix \ref{sec:metropolisstep} with sampling distribution
\begin{equation}
\label{samplingdisthomoskedastic}
\mathbb{P}(\alpha | \textbf{T}, \textbf{X}, \Theta_{-\alpha}) \propto \mathbb{P}(\textbf{T}, \textbf{X} |  \Theta_{-\alpha}) \Big(\prod_{y=1}^Y\mathbb{P}(\beta_{y} | \beta_{y - 1})\Big)\mathbb{P}(\beta_0)\mathbb{P}(\sigma_0^2)\mathbb{P}(\delta)\mathbb{P}(\nu)\mathbb{P}(\tau),
\end{equation}
where the likelihood and priors are as described in Section \ref{sec:homoskedasticmodel}.
Here, the parameter blocks in the Metropolis-within-Gibbs sampler are $\beta_y$ for each $y$, $\sigma_0^2$, $\delta$, $\nu$, and $\tau$.

\subsection{Heteroskedastic and Robust Heteroskedastic Model Parameters}
\label{sec:heterosample}

The set of model parameters in both the heteroskedastic and robust heteroskedastic models described in Sections \ref{sec:heteromodel} and \ref{sec:robustheteromodel} respectively is $\Theta = $\{$\beta_y$ \text{for all $y$ in our data}, $\gamma$, $\delta$, $\nu$, $\tau$\}. For this model, we set the hyperparameters to be $\mu_{\beta} = 0$, $\sigma^2_{\beta} = 100$, $\mu_{\gamma} = 0$, $\sigma^2_{\gamma} = 100$, $\lambda_0 = 1$, $a_\nu = 0$, $b_\nu = 50$, $a_\tau = -10$, and $b_\tau = 10$. As with the homoskedastic model, we choose our hyperpriors to be flat and uninformative. 

We note that for these models, the posterior cannot be written in closed form for any of the model parameters due to lack of conjugacy between the data likelihood $\mathbb{P}(\textbf{T}, \textbf{X} | \Theta)$ and the parameter priors. Therefore, we sample each parameter block $\alpha \in \Theta$ using a random walk Metropolis step according to the rules Section \ref{sec:metropolisstep} with sampling distribution
\begin{equation}
\label{samplingdisthetero}
\mathbb{P}(\alpha | \textbf{T}, \textbf{X}, \Theta_{-\alpha}) \propto \mathbb{P}(\textbf{T}, \textbf{X} |  \Theta_{-\alpha}) \Big(\prod_{y=1}^Y\mathbb{P}(\beta_{y} | \beta_{y - 1})\Big)\mathbb{P}(\beta_0)\mathbb{P}(\gamma)\mathbb{P}(\delta)\mathbb{P}(\nu)\mathbb{P}(\tau),
\end{equation}
where the appropriate likelihood and priors are used for each model type described in Sections \ref{sec:heteromodel} and \ref{sec:robustheteromodel}. Here, the parameter blocks in the Metropolis-within-Gibbs sampler are $\beta_y$ for each $y$, $\gamma$, $\delta$, $\nu$, and $\tau$.

\section{Model Portfolios }
\label{app:portfoliotablesall}

Here, we show the top ten companies in both the independent and correlated portfolios selected by each of our  models for the test years 2011 and 2012. The companies which exited in each portfolio are highlighted in bold and the exit probabilities and objective values are rounded to the nearest hundredth.



\begin{table}[h]
	\centering
	\caption{The top companies in our 2011 portfolio constructed with greedy optimization and the \textbf{homoskedastic independent} model.}
	\label{table:m0indepporfolios2011}
	\begin{tabular}{|c|c|c|}
		\hline
		\textbf{2011 Company} & \textbf{Highest funding round} & \textbf{Exit probability}\\	\hline
Sequent & B & 0.65\\   \hline
Syapse & C & 0.31\\   \hline
AirTouch Communications & A & 0.29\\   \hline
Whitetruffle & Seed & 0.29\\   \hline
Metromile & D & 0.29\\   \hline
\textbf{Funzio} & Acquisition & 0.22\\   \hline
\textbf{Bitzer Mobile} & Acquisition & 0.19\\   \hline
Axtria & C & 0.17\\   \hline
\textbf{Livestar} & Acquisition & 0.16\\   \hline
nprogress & Seed & 0.08\\   \hline
	\end{tabular}
\end{table}

\begin{table}[h]
	\centering
	\caption{The top companies in our 2011 portfolio constructed with greedy optimization and the \textbf{homoskedastic correlated} model.}
	\label{table:m0corrporfolios2011}
	\begin{tabular}{|c|c|c|c|c|}
		\hline
		\textbf{2011 Company} & \textbf{Highest funding round} & \textbf{Exit probability}&\textbf{ Objective value}\\	\hline
Sequent & B & 0.57 & 0.57 \\   \hline
AirTouch Communications & A & 0.35 & 0.71\\   \hline
Syapse & C & 0.34 & 0.79 \\   \hline
\textbf{Bitzer Mobile} & Acquisition & 0.3 & 0.83 \\   \hline
Whitetruffle & Seed & 0.4 & 0.85\\   \hline
\textbf{Funzio} & Acquisition & 0.29 & 0.87 \\   \hline
Metromile & D & 0.4 & 0.88 \\   \hline
nprogress & Seed & 0.17 & 0.89\\   \hline
\textbf{Livestar} & Acquisition & 0.2 & 0.9 \\   \hline
\textbf{Snapguide} & Acquisition & 0.14 & 0.91 \\   \hline
	\end{tabular}
\end{table}

\begin{table}[h]
	\centering
	\caption{The top companies in our 2012 portfolio constructed with greedy optimization and the \textbf{homoskedastic independent} model.}
	\label{table:m0indepporfolios2012}
	\begin{tabular}{|c|c|c|}
		\hline
		\textbf{2012 Company} & \textbf{Highest funding round} & \textbf{Exit probability}\\	\hline
Boomerang & Seed & 0.22\\   \hline
\textbf{Struq} & Acquisition & 0.21\\   \hline
Glossi  Inc & Seed & 0.21\\   \hline
Trv & C & 0.2\\   \hline
ScaleGrid & Seed & 0.07\\   \hline
UpGuard & B & 0.06\\   \hline
\textbf{Viewfinder} & Acquisition & 0.04\\   \hline
Mozio & Seed & 0.04\\   \hline
HighGround & A & 0.04\\   \hline
iLumen & Seed & 0.04\\   \hline
	\end{tabular}
\end{table}

\begin{table}[h]
	\centering
	\caption{The top companies in our 2012 portfolio constructed with greedy optimization and the \textbf{homoskedastic correlated} model.}
	\label{table:m0corrporfolios2012}
	\begin{tabular}{|c|c|c|c|c|}
		\hline
		\textbf{2012 Company} & \textbf{Highest funding round} & \textbf{Exit probability}&\textbf{ Objective value}\\	\hline
\textbf{Struq} & Acquisition & 0.32 & 0.32\\   \hline
Trv & C & 0.29 & 0.51\\   \hline
ScaleGrid & Seed & 0.25 & 0.63\\   \hline
Glossi  Inc & Seed & 0.31 & 0.68 \\   \hline
Boomerang & Seed & 0.3 & 0.72 \\   \hline
UpGuard & B & 0.17 & 0.74\\   \hline
\textbf{Viewfinder} & Acquisition & 0.1 & 0.75 \\   \hline
The Zebra & A & 0.12 & 0.76 \\   \hline
Bolster & Seed & 0.09 & 0.77 \\   \hline
Alicanto & A & 0.13 & 0.77\\   \hline
	\end{tabular}
\end{table}



\begin{table}[h]
	\centering
	\caption{The top companies in our 2011 portfolio constructed with greedy optimization and the \textbf{heteroskedastic independent} model.}
	\label{table:m1indepporfolios2011}
	\begin{tabular}{|c|c|c|}
		\hline
		\textbf{2011 Company} & \textbf{Highest funding round} & \textbf{Exit probability}\\	\hline
Metromile & D & 0.47\\   \hline
\textbf{Leaky} & Acquisition & 0.27\\   \hline
Syapse & C & 0.25\\   \hline
AirTouch Communications & A & 0.21\\   \hline
\textbf{Snapguide} & Acquisition & 0.2\\   \hline
\textbf{Fullscreen} & Acquisition & 0.13\\   \hline
Axtria & C & 0.1\\   \hline
\textbf{Livestar} & Acquisition & 0.08\\   \hline
\textbf{TappIn} & Acquisition & 0.07\\   \hline
BloomBoard & B & 0.03\\   \hline
	\end{tabular}
\end{table}

\begin{table}[h]
	\centering
	\caption{The top companies in our 2011 portfolio constructed with greedy optimization and the \textbf{heteroskedastic correlated} model.}
	\label{table:m1corrporfolios2011}
	\begin{tabular}{|c|c|c|c|c|}
		\hline
		\textbf{2011 Company} & \textbf{Highest funding round} & \textbf{Exit probability}&\textbf{ Objective value}\\	\hline
Metromile & D & 0.49 & 0.49\\   \hline
\textbf{Snapguide} & Acquisition & 0.27 & 0.63\\   \hline
Syapse & C & 0.28 & 0.72\\   \hline
Axtria & C & 0.41 & 0.78 \\   \hline
AirTouch Communications & A & 0.27 & 0.83\\   \hline
NextDocs & C & 0.25 & 0.86\\   \hline
Striiv & A & 0.11 & 0.88 \\   \hline
\textbf{Leaky} & Acquisition & 0.36 & 0.89 \\   \hline
\textbf{Bitzer Mobile} & Acquisition & 0.13 & 0.91 \\   \hline
AdexLink & Seed & 0.08 & 0.91\\   \hline
	\end{tabular}
\end{table}

\begin{table}[h]
	\centering
	\caption{The top companies in our 2012 portfolio constructed with greedy optimization and the \textbf{heteroskedastic independent} model.}
	\label{table:m1indepporfolios2012}
	\begin{tabular}{|c|c|c|}
		\hline
		\textbf{2012 Company} & \textbf{Highest funding round} & \textbf{Exit probability}\\	\hline
UpGuard & B & 0.56\\   \hline
Bolster & Seed & 0.53\\   \hline
Beam Dental & A & 0.34\\   \hline
\textbf{Struq} & Acquisition & 0.26\\   \hline
\textbf{Flutter} & Acquisition & 0.25\\   \hline
Glossi  Inc & Seed & 0.25\\   \hline
HighGround & A & 0.19\\   \hline
iLumen & Seed & 0.18\\   \hline
Camino Real & A & 0.18\\   \hline
\textbf{Newsle} & Acquisition & 0.17\\   \hline
	\end{tabular}
\end{table}

\begin{table}[h]
	\centering
	\caption{The top companies in our 2012 portfolio constructed with greedy optimization and the \textbf{heteroskedastic correlated} model.}
	\label{table:m1corrporfolios2012}
	\begin{tabular}{|c|c|c|c|c|}
		\hline
		\textbf{2012 Company} & \textbf{Highest funding round} & \textbf{Exit probability}&\textbf{ Objective value}\\	\hline
UpGuard & B & 0.55 & 0.55 \\   \hline
Bolster & Seed & 0.53 & 0.76 \\   \hline
Beam Dental & A & 0.4 & 0.83 \\   \hline
\textbf{Struq} & Acquisition & 0.3 & 0.88 \\   \hline
Glossi  Inc & Seed & 0.29 & 0.91 \\   \hline
\textbf{Flutter} & Acquisition & 0.29 & 0.93 \\   \hline
iLumen & Seed & 0.2 & 0.94 \\   \hline
Camino Real & A & 0.2 & 0.95 \\   \hline
ScaleGrid & Seed & 0.27 & 0.96 \\   \hline
HighGround & A & 0.2 & 0.96\\   \hline
	\end{tabular}
\end{table}



\begin{table}[h]
	\centering
	\caption{The top companies in our 2011 portfolio constructed with greedy optimization and the \textbf{robust heteroskedastic independent} model.}
	\label{table:m3indepporfolios2011}
	\begin{tabular}{|c|c|c|}
		\hline
		\textbf{2011 Company} & \textbf{Highest funding round} & \textbf{Exit probability}\\	\hline
\textbf{Funzio} & Acquisition & 0.55\\   \hline
AirTouch Communications & A & 0.48\\   \hline
\textbf{Bitzer Mobile} & Acquisition & 0.46\\   \hline
Sequent & B & 0.45\\   \hline
\textbf{SHIFT} & Acquisition & 0.43\\   \hline
\textbf{Longboard Media} & Acquisition & 0.41\\   \hline
\textbf{Livestar} & Acquisition & 0.37\\   \hline
\textbf{Jybe} & Acquisition & 0.37\\   \hline
Whitetruffle & Seed & 0.37\\   \hline
Adynxx & A & 0.32\\   \hline
	\end{tabular}
\end{table}

\begin{table}[h]
	\centering
	\caption{The top companies in our 2011 portfolio constructed with greedy optimization and the \textbf{robust heteroskedastic correlated} model.}
	\label{table:m3corrporfoliosapp2011}
	\begin{tabular}{|c|c|c|c|c|}
		\hline
		\textbf{2011 Company} & \textbf{Highest funding round} & \textbf{Exit probability}&\textbf{ Objective value}\\	\hline
                        \textbf{Funzio} & Acquisition & 0.54 & 0.54 \\   \hline
                        \textbf{Longboard Media} & Acquisition & 0.43 & 0.74\\   \hline
                        AirTouch Communications & A & 0.49 & 0.86 \\   \hline
                        Sequent & B & 0.47 & 0.92 \\   \hline
                        \textbf{Bitzer Mobile} & Acquisition & 0.47 & 0.95 \\   \hline
                        \textbf{SHIFT} & Acquisition & 0.44 & 0.96 \\   \hline
                        \textbf{Livestar} & Acquisition & 0.38 & 0.97 \\   \hline
                        \textbf{Jybe} & Acquisition & 0.4 & 0.98 \\   \hline
                        Whitetruffle & Seed & 0.4 & 0.99 \\   \hline
                        Adteractive & A & 0.27 & 0.99 \\   \hline
	\end{tabular}
\end{table}

\begin{table}[h]
	\centering
	\caption{The top companies in our 2012 portfolio constructed with greedy optimization and the \textbf{robust heteroskedastic independent} model.}
	\label{table:m3indepporfolios2012}
	\begin{tabular}{|c|c|c|}
		\hline
		\textbf{2012 Company} & \textbf{Highest funding round} & \textbf{Exit probability}\\	\hline
\textbf{Struq} & Acquisition & 0.8\\   \hline
Glossi  Inc & Seed & 0.79\\   \hline
\textbf{Metaresolver} & Acquisition & 0.74\\   \hline
Boomerang & Seed & 0.74\\   \hline
Trv & C & 0.7\\   \hline
Mozio & Seed & 0.31\\   \hline
Alicanto & A & 0.25\\   \hline
adRise & B & 0.21\\   \hline
\textbf{Runa} & Acquisition & 0.19\\   \hline
\textbf{SnappyTV} & Acquisition & 0.17\\   \hline
	\end{tabular}
\end{table}

\begin{table}[h]
	\centering
	\caption{The top companies in our 2012 portfolio constructed with greedy optimization and the \textbf{robust heteroskedastic correlated} model.}
	\label{table:m3corrporfoliosapp2012}
	\begin{tabular}{|c|c|c|c|c|}
		\hline
		\textbf{2012 Company} & \textbf{Highest funding round} & \textbf{Exit probability}&\textbf{ Objective value}\\	\hline
                        \textbf{Struq} & Acquisition & 0.76 & 0.76 \\   \hline
                        \textbf{Metaresolver} & Acquisition & 0.73 & 0.93 \\   \hline
                        Trv & C & 0.67 & 0.97 \\   \hline
                        Glossi  Inc & Seed & 0.75 & 0.99 \\   \hline
                        Boomerang & Seed & 0.71 & 0.99 \\   \hline
                        Alicanto & A & 0.3 & 1.0 \\   \hline
                        \textbf{SnappyTV} & Acquisition & 0.25 & 1.0 \\   \hline
                        AVOS Systems & A & 0.17 & 1.0 \\   \hline
                        \textbf{Adept Cloud} & Acquisition & 0.17 & 1.0 \\   \hline
                        \textbf{Runa} & Acquisition & 0.21 & 1.0 \\   \hline
	\end{tabular}
\end{table}

\end{APPENDICES}

\clearpage

\bibliographystyle{plainnat}
\bibliography{Winning_v03.bbl}

\end{document}